\documentclass[10pt, a4paper, conference, compsocconf]{IEEEtran}

\IEEEoverridecommandlockouts

\usepackage{url}
\usepackage{graphicx}
\usepackage{listings}
\usepackage{paralist}
\usepackage{multirow}
\usepackage[utf8]{inputenc}
\usepackage[cmex10]{amsmath}
\usepackage{amssymb}
\usepackage{xspace}
\usepackage[numbers,sort&compress]{natbib}
\usepackage{lscape}
\usepackage[table,xcdraw]{xcolor}
\usepackage{verbatim} % comment
\usepackage{amsmath}
\usepackage{booktabs} % style for plain tables
\usepackage{colortbl} % for colouring tables
\usepackage{blindtext}
\usepackage{hyperref}
\usepackage{tikz}
\usepackage{framed} % shaded boxes
\definecolor{shadecolor}{rgb}{0.85,0.85,0.85}
\definecolor{right} {HTML}{3F3F3F}

\usepackage{subcaption}

\usepackage{comment}

\usepackage[normalem]{ulem} % for \sout
\usepackage{xcolor}
\AtBeginDocument{%
  \providecommand\BibTeX{{%
    \normalfont B\kern-0.5em{\scshape i\kern-0.25em b}\kern-0.8em\TeX}}}

\usepackage{ifthen}
\usepackage{amssymb}
\newboolean{showcomments}
\setboolean{showcomments}{false} % toggle to show or hide comments
\ifthenelse{\boolean{showcomments}}
  {\newcommand{\nb}[2]{
    \fcolorbox{gray}{yellow}{\bfseries\sffamily\scriptsize#1}
    {\sf\small$\blacktriangleright$\textit{#2}$\blacktriangleleft$}
   }
   
  }
  {\newcommand{\nb}[2]{}
   
  }
\newcommand{\rdl}[1]{\nb{Rogerio}{\color{red}\footnotesize #1}}

\newcommand{\fcf}[1]{\nb{Fabiano}{\color{red}\footnotesize #1}}

\definecolor{shadecolor}{rgb}{0.8,0.8,0.8}
\definecolor{right} {HTML}{3F3F3F}

\newcommand{\ie}{\textit{i.e.}\xspace}
\newcommand{\eg}{\textit{e.g.}\xspace}

\usepackage{color}
\definecolor{darkgreen}{cmyk}{1,0,1,0}
\newcommand{\fabiano}[1]{{\color{black} #1}}
\newcommand{\bento}[1]{{\color{black} #1}}

\newcommand{\rogerio}[1]{{\color{black} #1}}

\newcommand{\important}[1]{{\color{red} [IMPORTANT: #1]}}

\newcommand{\phoneAdapter}{\emph{PhoneAdapter}\xspace}

\newcommand{\newPhoneAdapter}{\emph{$\mu$PhoneAdapter}\xspace}

\pgfarrowsdeclare{bad latex}{bad latex}
{
  \pgfarrowsleftextend{-1\pgflinewidth}
  \pgfarrowsrightextend{9\pgflinewidth}
}
{
  \pgfpathmoveto{\pgfpoint{9\pgflinewidth}{0pt}}
  \pgfpathcurveto
  {\pgfpoint{6.3333\pgflinewidth}{.5\pgflinewidth}}
  {\pgfpoint{2\pgflinewidth}{2\pgflinewidth}}
  {\pgfpoint{-1\pgflinewidth}{3.75\pgflinewidth}}
  \pgfpathlineto{\pgfpoint{-1\pgflinewidth}{-3.75\pgflinewidth}}
  \pgfpathcurveto
  {\pgfpoint{2\pgflinewidth}{-2\pgflinewidth}}
  {\pgfpoint{6.3333\pgflinewidth}{-.5\pgflinewidth}}
  {\pgfpoint{9\pgflinewidth}{0pt}}
  \pgfusepathqfill
}

\pgfarrowsdeclare{bad to}{bad to}
{
  \pgfarrowsleftextend{-2\pgflinewidth}
  \pgfarrowsrightextend{\pgflinewidth}
}
{
  \pgfsetlinewidth{0.8\pgflinewidth}
  \pgfsetdash{}{0pt}
  \pgfsetroundcap
  \pgfsetroundjoin
  \pgfpathmoveto{\pgfpoint{-3\pgflinewidth}{4\pgflinewidth}}
  \pgfpathcurveto
  {\pgfpoint{-2.75\pgflinewidth}{2.5\pgflinewidth}}
  {\pgfpoint{0pt}{0.25\pgflinewidth}}
  {\pgfpoint{0.75\pgflinewidth}{0pt}}
  \pgfpathcurveto
  {\pgfpoint{0pt}{-0.25\pgflinewidth}}
  {\pgfpoint{-2.75\pgflinewidth}{-2.5\pgflinewidth}}
  {\pgfpoint{-3\pgflinewidth}{-4\pgflinewidth}}
  \pgfusepathqstroke
}

\begin{document}
%\conferenceinfo{WOODSTOCK}{'97 El Paso, Texas USA}
%\CopyrightYear{2007} % Allows default copyright year (20XX) to be over-ridden - IF NEED BE.
%\crdata{0-12345-67-8/90/01}  % Allows default copyright data (0-89791-88-6/97/05) to be over-ridden - IF NEED BE.
% --- End of Author Metadata ---

\title{Micro-controllers: Promoting Structurally Flexible Controllers in Self-Adaptive Software Systems
}

\author{
\IEEEauthorblockN{Bento R. Siqueira, Fabiano C. Ferrari}
\IEEEauthorblockA{\textit{Federal University of S\~ao Carlos}\\
%S\~ao Carlos, SP, Brazil \\
bento.siqueira, fcferrari @ufscar.br}
\and
\IEEEauthorblockN{Thomas Vogel}
\IEEEauthorblockA{\textit{Humboldt-Universität zu Berlin}\\
%Berlin, Germany \\
thomas.vogel@cs.hu-berlin.de}
\and
\IEEEauthorblockN{Rog\'erio de Lemos}
\IEEEauthorblockA{\textit{School of Computing,} 
\textit{University of Kent}\\
%Canterbury, Kent, United Kingdom \\
r.delemos@kent.ac.uk }
}

\maketitle

\begin{abstract}
  To promote structurally flexible controllers in self-adaptive software systems, this paper proposes the use of micro-controllers. 
Instead of generic monolithic controllers, like Rainbow, we advocate the use of service-specific micro-controllers which can be based on microservices.
Although traditional generic controllers can be configured parametrically according to system needs, their use and reuse are nevertheless restrictive because of the wide range of services expected from the different stages of the feedback control loop. 
The solution being advocated is to have structurally flexible controllers that can be composed from micro-controllers. 
Controlling the architectural configuration of these micro-controllers is a meta-controller that is able to configure the controller according to the services required for controlling the target system.
The feasibility of the proposed approach of using micro-controllers at the level of the controller is demonstrated in the context of the PhoneAdapter case study in which micro-controllers are configured at run-time depending on changes affecting the system or its environment. 

% for EasyChair system

% To promote structurally flexible controllers in self-adaptive software systems, this paper proposes the use of micro-controllers. Instead of generic monolithic controllers, like Rainbow, we advocate the use of service-specific micro-controllers which can be based on microservices. Although traditional generic controllers can be configured parametrically according to system needs, their use and reuse are nevertheless restrictive because of the wide range of services expected from the different stages of the feedback control loop. The solution being advocated is to have structurally flexible controllers that can be composed from micro-controllers. Controlling the architectural configuration of these micro-controllers is a meta-controller that is able to configure the controller according to the services required for controlling the target system. The feasibility of the proposed approach of using micro-controllers at the level of the controller is demonstrated in the context of the PhoneAdapter case study in which micro-controllers are configured at run-time depending on changes affecting the system or its environment. 
\end{abstract}


\begin{sloppypar}

%-----------------------------------------

\begin{comment}
\important{NEW DEADLINES DUE TO COVID-19 CRISIS:

\begin{itemize}

    \item \textbf{Size limit: 10 pages including references}

    \item ACSOS 2020 - 1st IEEE International Conference on Autonomic Computing and Self-Organizing Systems
    
    \item \url{https://conf.researchr.org/home/acsos-2020}
    
    \item Fri 8 May 2020: Research Papers -- Abstract Submission
    
    \item Fri 15 May 2020: Research Papers -- Paper Submission
    
    \item Mon 8 Jun 2020: Notification to Authors
    
    \item Wed 8 Jul 2020: Camera Ready

\end{itemize}
}
\end{comment}

\section{Introduction} 
\label{sec:introduction}

Rainbow~\cite{garlan2004} is one of the few examples of a \textit{generic} controller for self-adaptive software systems that has been used in several application domains~\cite{yuan2013,schmerl2014} and by different researchers~\cite{camara2013}. 
%Many of the existing controllers cannot be reused across different applications and by different developers. 
In general, a key factor restricting their reuse is that controllers are intrinsically dependent on the target system (the software system to be controlled) beyond the probes and effectors wiring the controller to the target system. 
% This coupling is not restricted to probes and effectors that connect the controller to the target system, but it is mainly related to the inherent dependencies between the controller and the target system. 
Particularly, a controller has to be tailored to the target system-specific intricacies and complexities, such as the goals and configuration space of the system and how the system behaves in different configurations and contextual states with respect to the goals.
% So, a major factor that affects the reuse of controllers is the need to tailor controllers to such intricacies and complexities of the target system.
If a controller is not able to appropriately abstract from such intricacies and complexities, the coupling between the controller and target system becomes~unmanageable to~the point that the prospect of reusing a controller becomes minimal.

To mitigate this issue, Rainbow raises its abstraction of a target system to the software architecture, which successfully reduces the coupling between the controller and system~\cite{garlan2004}. 
However, the wide context and role of the software of a target system may also impose a wide range of services that are expected from the different stages of a controller.
For instance, using a MAPE-K controller~\cite{kephart2003}, a safety-critical target system may need an analysis stage based on formal model checking while another system may only need an architectural analysis.
% an analysis stage using formal model checking may be needed for safety-critical systems while some kind of architectural analysis could be sufficient for a normal system. 
Adopting the Rainbow approach to develop a \textit{truly generic} controller that is able to provide such a wide range of services is quite challenging, and perhaps counter-productive since the unit of reuse would be a monolithic controller that is only parametrically configurable, which is the case in most existing controllers~\cite{krupitzer2016}.
In contrast, a controller whose provided services are composed according to the target system's actual needs may achieve 
% \fcf{"achieves" or "would achieve"?}
a higher flexibility and variability than a monolithic controller---similarly to a target system that can achieve a higher flexibility and variability when being adapted structurally and not only parametrically~\cite{mcKinley2004}. 
By adopting such a modular and structurally flexible approach to build a truly generic controller, the individual services of a controller \rogerio{should become structurally independent from each other. }

%\ap{[Rev. 1] Here we can first introduce the idea of micro-controllers as a key concept of our approach (as done in the Abstract and in Section~\ref{sec:aprroach}), and then we suggest microservices as possible realisation of the approach due to their intrinsic concepts and associated technology.}

\rogerio{To achieve structurally flexible controllers,
%Consequently, 
we propose the use of \textit{micro-controllers}
%, each being a microservice 
for implementing closed or open control loops.
These micro-controllers, which could be deployed as microservices, are able} to provide fine-granular functionality for self-adaptation, while choreography 
%\ap{[Rev.1] Here we mention "choreography". The Reviewer mentioned "orchestration" of microservices. I guess the Reviewer is wrong.}
of micro-controllers implements the overall control of the target system.
The choreography may typically follow a common %\ap{[Rev.1] replace "simple" with "common"?}
control flow according to the stages of a controller (\eg, MAPE) while data is exchanged between micro-controllers using a shared service maintaining the knowledge (\eg, models of the target system). 
The latter allows micro-controllers to be stateless. % if needed.
Essentially, we propose the synthesis of a controller from a collection of micro-controllers.

In this context, the main idea being promoted in this paper is that controllers do not need to be structurally static at development-time or run-time. 
In other words, controllers can both be the locus of change and the locus of adaptation. 
Therefore, a controller as an ensemble of micro-controllers can be composed and configured at development-time by developers or dynamically recomposed and reconfigured at run-time by a meta-controller based on the needs of a specific target system and the adaptation goals. 
%In both cases, such an approach promotes reuse of existing micro-controllers since the required needs of a target system and the adaptation goals are addressed by the structurally flexible ensemble of micro-controllers. 
A key challenge is to compose an ensemble of micro-controllers that is able to control a given target system (which would satisfy the adaptation goals) and to satisfy the system's specific needs.

Thus, the contribution of this paper is a novel approach to develop controllers by composing and configuring micro-con\-trollers 
%to a controller 
based on the adaptation goals and the needs of a given target system. 
The benefits of the approach are twofold. 
First, it leverages controllers being the locus of adaptation \textit{and} the locus of change as the micro-controllers provide the \textit{structural flexibility} required to recompose and reconfigure a controller. 
Second, it leverages \textit{reuse} by considering micro-controllers as fine-granular units that provide generic services to controllers.
We motivate, illustrate, and qualitatively evaluate our approach with the \textit{PhoneAdapter} case study~\cite{sama2008}, particularly focusing on the flexibility of~controllers.

%\todo{revise for a complete submission}
The rest of the paper is organised as follows.
%We provide the background of our work in Sec.~\ref{sec:background}. 
In Section~\ref{sec:caseStudy}, we introduce the case study.
We present the approach in Section~\ref{sec:aprroach},
demonstrate its use with the case study in Section~\ref{sec:demonstration},
% We demonstrate the approach using the case study in Section~\ref{sec:demonstration}.
and evaluate and discuss \rogerio{its contribution} in Section~\ref{sec:demonstration}. 
%Finally,
We discuss related work in Section~\ref{sec:relatedWork} and conclude 
% the paper 
in~Section~\ref{sec:conclusions}.

%\section{Background} 
%\label{sec:background}
%\input{background}

\section{Case Study} 
\label{sec:caseStudy}

\rogerio{To explain the design of controllers that are based on micro-controllers, we have adopted  the \emph{PhoneAdapter} case study that is described in the following.}
%, for self-adaptive software systems.
The \emph{PhoneAdapter} was originally proposed by the community of context-aware adaptive applications for exemplifying the definition of adaptation rules that depend on the system context~\citep{sama2008}.
The \emph{PhoneAdapter} is a mobile application for Android devices that allows users to set up system profiles and rules, corresponding to different contexts and adaptation rules, in order to adapt Android features.
%\rogerio{Context-aware applications, such as, the \textit{PhoneAdapter}, should be made adaptable to context changes.}
%Context-aware applications such as the \textit{PhoneAdapter} are intensely context-aware and adaptive as they continually adapt to context changes.
For that, the \textit{PhoneAdapter} relies on two components:
a \emph{context manager} that collects and maintains context information, and
an \emph{adaptation manager} that adapts the target system according to the context information and a set of adaptation~rules. \rogerio{The motivation for choosing this case study is its simplicity, which allows us to show how micro-controllers can be used to promote structurally flexible controllers following an hierarchical control pattern (cf.~\cite{weyns2013})}.
%\fcf{here we are emphasising "reuse"... double-check if we properly address it along the paper}
%of~controllers.

%\ap{Here (or perhaps in another place) we can briefly address the issues of Reviewer 3 that regards decentralised control.} \fcf{"On Patterns for Decentralised Control in Self-Adaptive Systems". We can refer to a paper by \citet{Weyns:2013:patterns}. That paper presents varied types of controllers and we can point to the type of controller that is characterised in our paper.}

\begin{figure}[!ht]
    \centering
    \includegraphics[width=0.5\columnwidth]{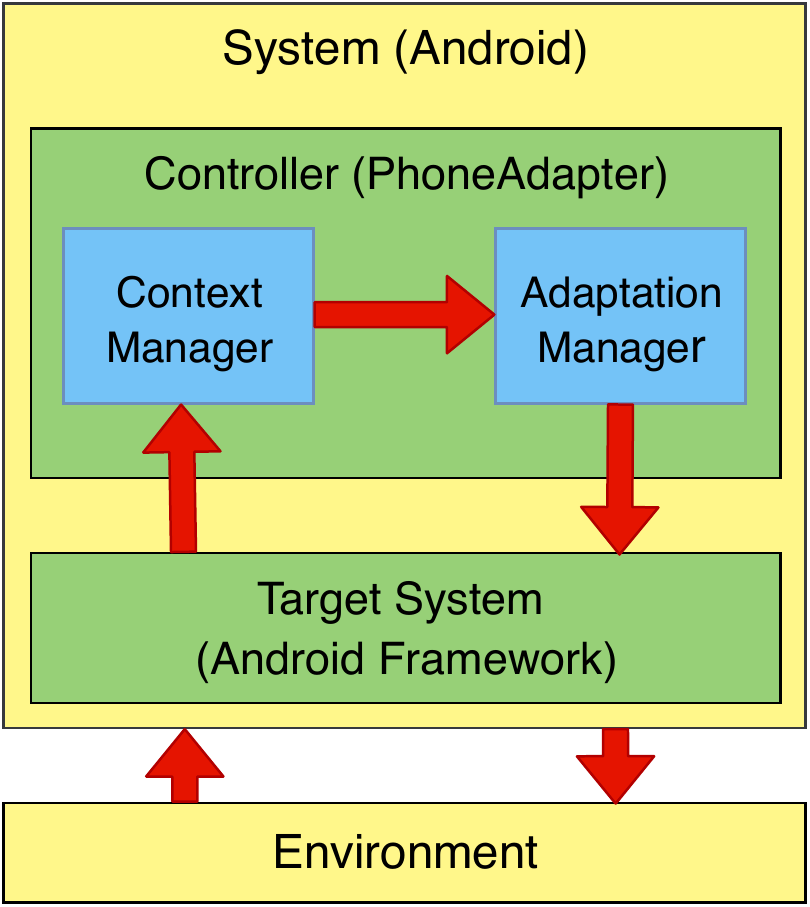}
    %\vspace{-.3cm}
    \caption{Android as a self-adaptive software system.
%    \fcf{Explain the notation and respective semantics (e.g. a blue rectangle represents a "component" of PhoneAdapter)... or should we leave this for the next sections?}
    }
    \label{fig:phoneAdapterArchitecture}
    \vspace{-1em}
\end{figure}

Figure~\ref{fig:phoneAdapterArchitecture} shows an overview of the original version of the \textit{PhoneAdapter}~\citep{sama2008} 
%as an Android application 
that turns \textsf{Android} into a self-adaptive system. 
The \textsf{PhoneAdapter} element is the controller, whereas the \textsf{Android Framework}
%\textsf{Android OS} 
is the target system.
% The system consists of the target system (\textsf{Android Framework}) and the controller (\textsf{PhoneAdapter}).
The controller contains two components: \textsf{ContextManager} and \textsf{AdaptationManager}.
%, plus \textsf{Knowledge}
%\brs{\bento{These Knowledge will be removed from the original version.}}.
The \textsf{ContextManager} component is responsible for identifying the context of the mobile device depending on the values of sensors provided by the 
\textsf{Android Framework}.
%\textsf{Android OS}.
The \textsf{AdaptationManager} component, on the other hand, is responsible for controlling the features of the \textsf{Android Framework} (\eg, ringtone volume) through effectors
%---also known as \emph{effectors} %\rdl{in the text we use interchangeably effectors and actuators, my suggestion is to change everything to effectors} \bento{done.}--- 
depending on the identified context.
% The original case study assumed that users do not have any direct interaction with the operation of the \textsf{PhoneAdapter} application.
The original case study assumes no interactions between users and the \textsf{PhoneAdapter} at run-time. Thus, there is no direct interaction between the mobile device's environment and the~\textsf{PhoneAdapter}.

\begin{figure*}[!ht]
    \centering
    \begin{subfigure}[h]{0.4\textwidth}
        \includegraphics[width=0.9\textwidth]{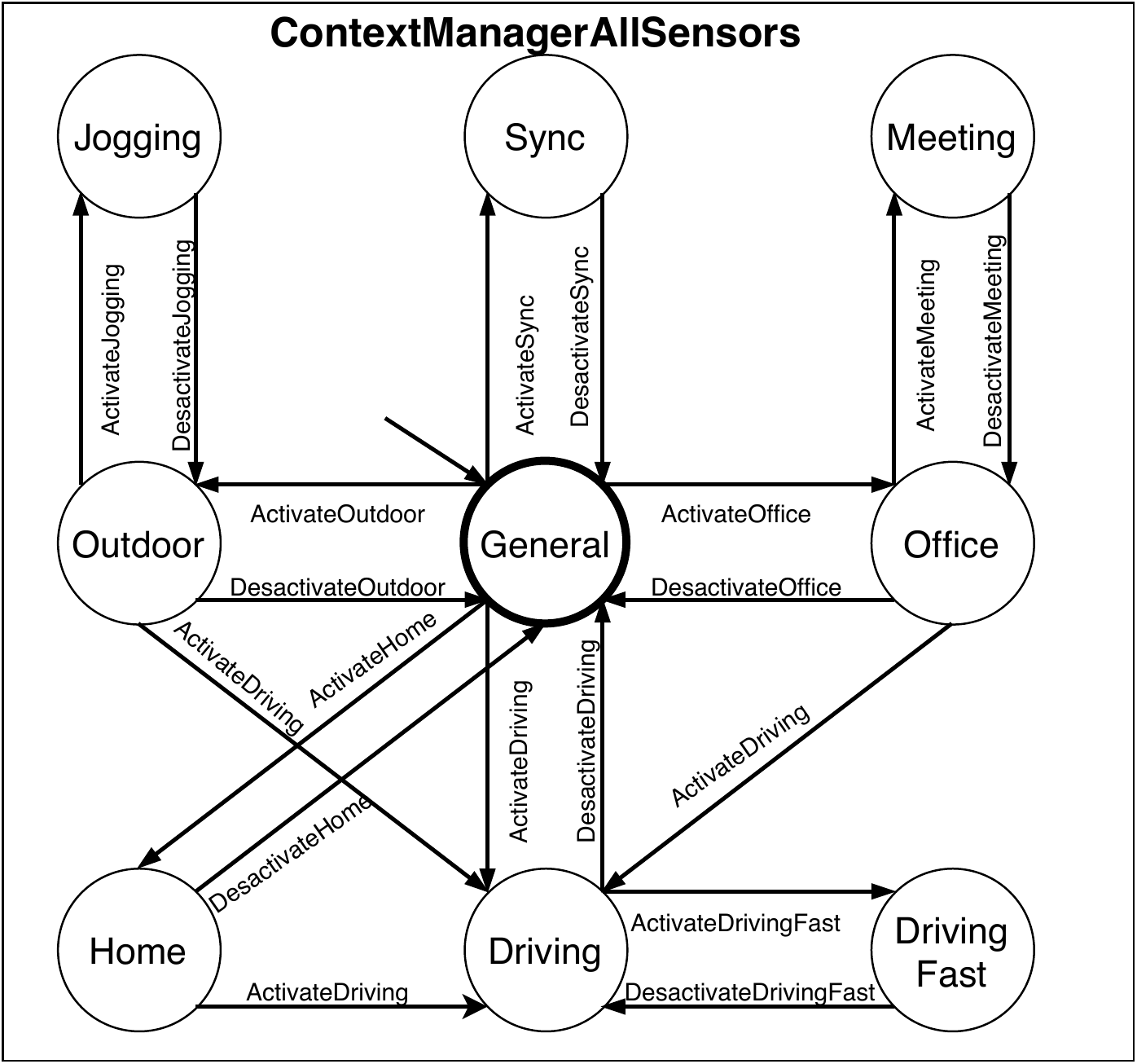}\hfill
        \subcaption{Original A-FSM~\citep{sama2008}.\label{fig:ContextManagerAllSensors}}
    \end{subfigure}
    \begin{subfigure}[h]{0.4\textwidth}
        \includegraphics[width=0.9\textwidth]{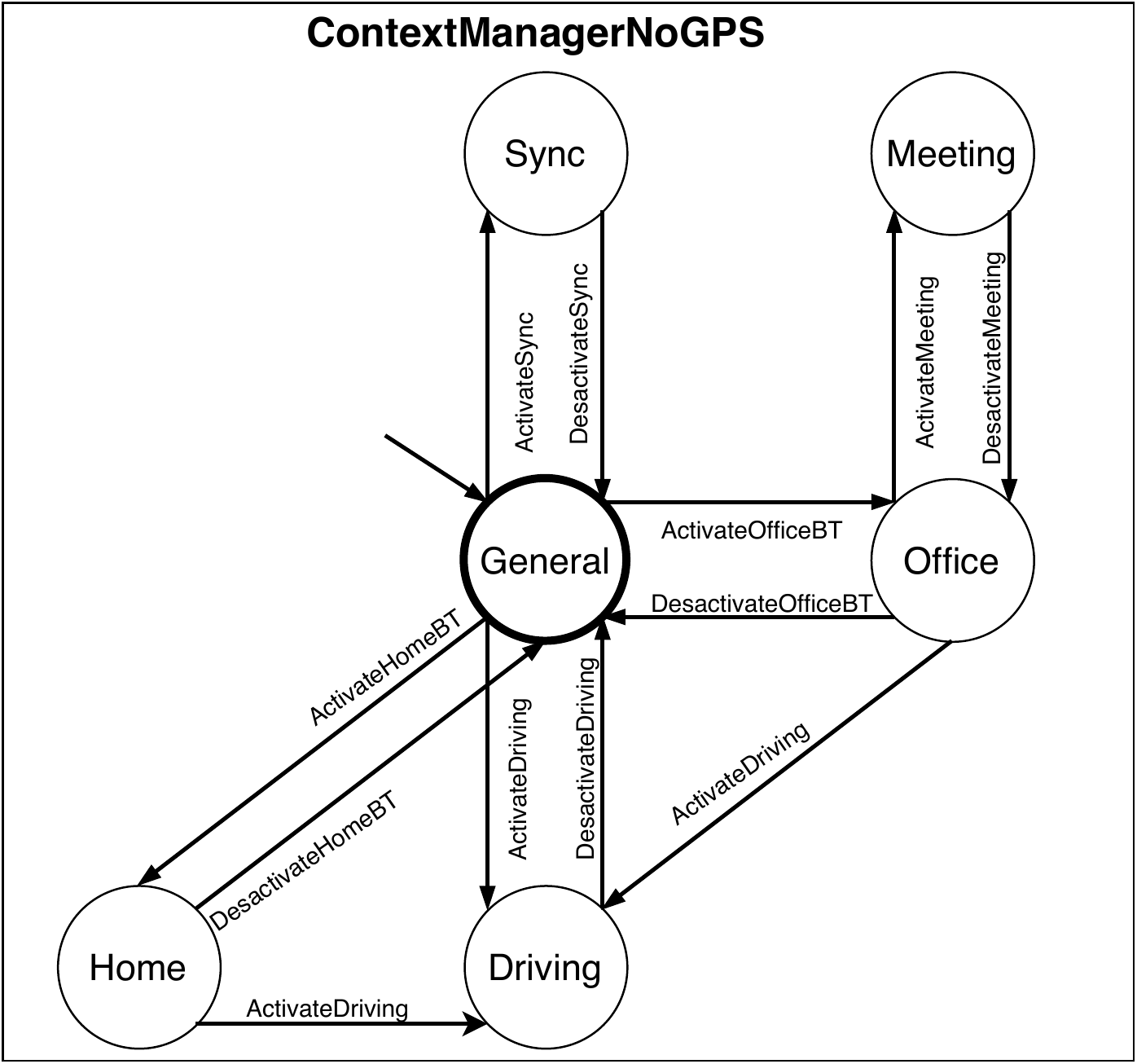}\hfill
        \subcaption{A-FSM for GPS sensor failure.\label{fig:ContextManagerNoGPS}}
    \end{subfigure}\\
    
    %\begin{subfigure}[h]{0.4\textwidth}
    %    \includegraphics[width=0.9\textwidth]{figures/ContextManagerNoBluetooth.pdf}\hfill
    %    \subcaption{A-FSM for Bluetooth sensor failure.\label{fig:ContextManagerNoBluetooth}}
    %\end{subfigure}
    %\begin{subfigure}[h]{0.4\textwidth}
    %    \includegraphics[width=0.9\textwidth]{figures/ContextManagerNoGPSNoBluetooth.pdf}\hfill
    %    \subcaption{A-FSM for GPS and Bluetooth sensor failures.\label{fig:ContextManagerNoGPSNoBluetooth}}
    %\end{subfigure}
    
    %\vspace{-.3cm}
    \caption{Two contextual spaces based on the availability of sensors, each defined by an \textit{adaptation finite state machine} (A-FSM).
    %\note{The link to editable figure of FSMs is \url{https://drive.google.com/file/d/1kYVEjZ4VYvNgUhbBq7Yg-8OIribwEIrZ/view?usp=sharing}.}
    %\note{The link to editable tables of rules \url{https://docs.google.com/spreadsheets/d/1-vBEzH5NYgXUQTYqu4tggbkwUT7vbLeI7NP3gtgrMIU/edit?usp=sharing}}
    }
    \label{fig:ContextManagers}
\end{figure*}

\begin{table*}[!ht]
    \centering
    \caption{\bento{Contextual} rules for the \textsf{ContextManagerAllSensors} A-FSM.}
    
			\centering
			%\small

            \vspace{-.2cm}

	        \setlength{\tabcolsep}{1pt}
            %\scriptsize   
            % \small
            
            \fontencoding{T1}
            %\fontfamily{\ttdefault}
            \fontfamily{\rmdefault}
            \fontseries{m}
            \fontshape{n}
            \fontsize{7.7}{6.7}
            \selectfont   
            
            %\vspace{-0.2cm}
    \label{tabAdaptationRulesContextManagerAllSensors}

%\begin{tabular}{cp{3cm}p{2.1cm}p{1.5cm}p{6.4cm}p{1.2cm}p{1.5cm}p{1cm}} 

\begin{tabular}{cp{2.8cm}p{3.5cm}p{1.5cm}p{7.5cm}cc} 

\toprule

\rowcolor{lightgray!50}
 \multicolumn{5}{c}{}  & \multicolumn{2}{c}{\textbf{Output}}  \\ \rowcolor
 {lightgray!50}
\textbf{Id} & \textbf{Rule Name}              & \textbf{Current State}      & \textbf{New State}  & \textbf{Full Predicate} & \textbf{Volume} & \textbf{Vibration}  \\ \midrule

a  & ActivateOutdoor        & General            & Outdoor     & GPS.isValid() \&\& !GPS.location()=home \&\& !GPS.location=office  & 100 & OFF         \\
\rowcolor{lightgray!50}
b  & DesactivateOutdoor     & Outdoor            & General     & !ActivateOutdoor & 50 & OFF         \\

c  & ActivateJogging        & Outdoor            & Jogging     & GPS.isValid() \&\& GPS.speed() \textgreater 5 & 25  & OFF   \\
\rowcolor{lightgray!50}
d  & DesactivateJogging     & Jogging            & Outdoor     & !ActivateJogging                              & 100 & OFF   \\

e  & ActivateDriving        & General, Home, Office, Outdoor & Driving & BT=car\_handsfree                     & 75  & OFF   \\
\rowcolor{lightgray!50}
f  & DesactivateDriving   & Driving & General & !ActivateDriving 																				   & 50 & OFF   \\

g  & ActivateDrivingFast  & Driving & DrivingFast & GPS.isValid() \&\& GPS.speed() \textgreater 70 												   & 0 & OFF  \\
\rowcolor{lightgray!50}
h  & DesactivateDrivingFast& DrivingFast & Driving & !ActivateDrivingFast 																		   & 75 & OFF  \\

i  & ActivateHome          & General     & Home    & BT=home\_pc || (GPS.isValid() \&\& GPS.location()=home) 									   & 100 & OFF  \\
\rowcolor{lightgray!50}
j & DesactivateHome        & Home        					& General     & !ActivateHome														   & 50 & OFF  \\

k & ActivateOffice         & General                        & Office      & BT=office\_pc || (GPS.isValid() \&\& GPS.location()=office)            & 0   & ON   \\
\rowcolor{lightgray!50}
l & DesactivateOffice      & Office                         & General     & !ActivateOffice                                                        & 50  & OFF  \\

m & ActivateMeeting        & Office                         & Meeting     & Time \textgreater{}= meeting\_start \&\& BT.count() \textgreater{}= 3  & 0   & OFF  \\
\rowcolor{lightgray!50}
n & DesactivateMeeting     & Meeting                        & Office      & Time \textgreater{}= meeting\_end                                      & 0   & ON   \\

o & ActivateSync           & General                        & Sync        & BT=home\_pc || BT=office\_pc                                           & 100 & OFF  \\
\rowcolor{lightgray!50}
p & DesactivateSync        & Sync                           & General     & !ActivateSync                                                          & 50  & OFF  \\ \bottomrule 

\end{tabular}
\end{table*}

\begin{table*}[!ht]
    \centering
    \caption{\bento{Contextual} rules for the \textsf{ContexManagerNoGPS} A-FSM.}
    
			\centering
			%\small

            \vspace{-.2cm}

	        \setlength{\tabcolsep}{1pt}
            %\scriptsize   
            % \small
            
            \fontencoding{T1}
            %\fontfamily{\ttdefault}
            \fontfamily{\rmdefault}
            \fontseries{m}
            \fontshape{n}
            \fontsize{7.7}{6.7}
            \selectfont   
            
            %\vspace{-0.2cm}
            \label{tabAdaptationRulesContextManagerNoGPS}
    
\begin{tabular}{cp{2.8cm}p{3.5cm}p{1.5cm}p{7.5cm}cc} 

\toprule

\rowcolor{lightgray!50}
 \multicolumn{5}{c}{}  & \multicolumn{2}{c}{\textbf{Output}}  \\ \rowcolor
 {lightgray!50}
\textbf{Id} & \textbf{Rule Name}              & \textbf{Current State}      & \textbf{New State}  & \textbf{Full Predicate} & \textbf{Volume} & \textbf{Vibration}  \\ \midrule
  
e & ActivateDriving        & General, Home, Office, Outdoor & Driving     & BT=car\_handsfree                                                     & 75     & OFF               \\
\rowcolor{lightgray!50}
f & DesactivateDriving     & Driving                        & General     & !ActivateDriving                                                      & 50     & OFF               \\

i & ActivateHomeBT         & General                        & Home        & BT=home\_pc                                                           & 100    & OFF               \\
\rowcolor{lightgray!50}
j & DesactivateHomeBT      & Home                           & General     & !ActivateHomeBT                                                       & 50     & OFF               \\

k & ActivateOfficeBT       & General                        & Office      & BT=office\_pc                                                         & 0      & ON                \\
\rowcolor{lightgray!50}
l & DesactivateOfficeBT    & Office                         & General     & !ActivateOfficeBT                                                     & 50     & OFF               \\

m & ActivateMeeting        & Office                         & Meeting     & Time \textgreater{}= meeting\_start \&\& BT.count() \textgreater{}= 3 & 0      & OFF               \\
\rowcolor{lightgray!50}
n & DesactivateMeeting     & Meeting                        & Office      & Time \textgreater{}= meeting\_end                                     & 0      & ON                \\

o & ActivateSync           & General                        & Sync        & BT=home\_pc || BT=office\_pc                                          & 100    & OFF               \\
\rowcolor{lightgray!50}
p & DesactivateSync        & Sync                           & General     & !ActivateSync                                                         & 50     & OFF             \\ \bottomrule

\end{tabular}
\end{table*}

In this paper, we assume that individual sensors and effectors may fail arbitrarily, and because of that, the number of possible configurations of the target system becomes combinatorial. Considering $n$ sensors and $m$ effectors, each being either operational or failing, $2^{n+m}$ different configurations exist for the target system.
To deal with this combinatorial explosion, we consider that specific \textsf{ContextManagers} and \textsf{AdaptationManagers} can be generated depending on the availability of sensors and effectors, respectively.

The \emph{PhoneAdapter} \rogerio{includes} two sensors, GPS and Bluetooth, and a probe to the calendar application. % while we treat this probe as a sensor.
These sensors and the probe can be accessed through the \textsf{Android Framework} and they allow the \textsf{ContextManager} to identify the different contexts of the mobile device.
The state machine---originally called \emph{Adaptation Finite-State Machine (A-FSM)}~\cite{sama2008}---shown in Figure~\ref{fig:ContextManagerAllSensors} consists of nine possible contextual states
% \fcf{In a footnote, should we say that this state machine depends on the actual set of adaptation rules. For consistency sake, we kept it is as originally modelled by~\citet{sama2008}.} Thomas: It is ok as it is.
depending on the values of the sensors and probe, with the \textsf{General} state being the initial state.
The transitions between the states are defined in terms of contextual rules that the \textsf{ContextManager} uses to identify context switches based on changing values of the sensors and probe~\citep{sama2008}. 
The contextual rules of Figure~\ref{fig:ContextManagerAllSensors} are presented in Table~\ref{tabAdaptationRulesContextManagerAllSensors}.
%\tv{Adaptation rules change the state of the target system rather than that of the context! We should be clear that based on sensor values and calendar entries, the rules change the state of the system according to the contextual state (location due to GPS, connectivity due to Bluetooth, availability due to calendar entries).}
%\rdl{the original paper called them adaptation rules, but to avoid any confusion we should use just rules}
%\tv{Shall the state machine transitions include triggers and conditions to apply a rule? Currently, the rule seems just to be an action without defining when the action is triggered and under which conditions it is executed (cf. event-condition-action rules).}
%\fcf{Explain the complete FSM}\rdl{there is no need}
Thus, nine different contexts can be identified based on the sensors and probe.
However, sensors may fail, which may result in a fewer number of identifiable contexts and thus in different contextual spaces depending on which and how many sensors fail at the same time.
%\brs{We mentioned that sensors may fail, however we just put examples of alternatives adaptation managers. Even tough we have already put in the figure.}
For example, if the GPS sensor fails, the \textsf{ContextManager} is not able to identify the \textsf{Jogging}, \textsf{Outdoor}, and \textsf{DrivingFast} contextual states, as shown in Figure~\ref{fig:ContextManagerNoGPS}. Table~\ref{tabAdaptationRulesContextManagerNoGPS} presents the contextual rules for Figure~\ref{fig:ContextManagerNoGPS}.
%\tv{Table I and II: their captions mention ``Adaptation rules'' not context rules as the text above; The tables show also some output on volume and vibration. This sounds like the adaptation that occurs, which however does not make sense for contextual rules!? The text below mentions that the table shows adaptation rules.}
%\ap{REMOVE THE URL}}
%\fabiano{\footnote{A complete set of alternative A-FSMs for the \phoneAdapter ---\ap{include tables as well}and associated sets of adaptation rules--- is available online at~\url{https://bit.ly/3buxWxF}.} 

%In the case that the \textsf{Bluetooth} sensor fails, the \textsf{Driving}, \textsf{DrivingFast}, and \textsf{Sync} contexts cannot be identified as shown in Figure~\ref{fig:ContextManagerNoBluetooth}.
%If both sensors, \textsf{Bluetooth} and \textsf{GPS} fail at the same time, the \textsf{ContextManager} is able to identify only two contexts: whether the user is in a meeting (\textsf{Meeting}) or not (\textsf{General}) (see Figure~\ref{fig:ContextManagerNoGPSNoBluetooth}).

%\fcf{Briefly explain the variations of the FSM}
% \fcf{Should we upfront emphasize the need for alternative ContextManagers?}\rdl{leave for the next section - but not yet sure how to approach this}

%\ap{Double-check the description of the AdaptationManager... see ActionPoint at the beggining of the section.}
Concerning the effectors, the \textsf{PhoneAdapter} controls different features of the \textsf{Android Framework}, such as the ringtone volume, vibration, backlight intensity, speaker volume, and divert phone calls.\footnote{For the sake of simplicity, the rules shown in Tables~\ref{tabAdaptationRulesContextManagerAllSensors} and~\ref{tabAdaptationRulesContextManagerNoGPS} only consider settings related to volume and vibration.}
For example, in the \textsf{Meeting} context, the \textsf{AdaptationManager} should mute the ringtone and disable vibration
(see rule \emph{m} in Tables~\ref{tabAdaptationRulesContextManagerAllSensors} and~\ref{tabAdaptationRulesContextManagerNoGPS}), 
while in the \textsf{Jogging} context, it should decrease
%increase the backlight intensity and 
speaker volume and deactivate vibration (see rule \emph{c} in Table~\ref{tabAdaptationRulesContextManagerAllSensors}). 
However, since individual effectors may fail, the adaptation rules have to be modified according to the availability of effectors.
Similarly to the state machines defining the identification of different contexts by the \textsf{ContextManager} (Figure~\ref{fig:ContextManagers}), different state machines defining the adaptation behaviour (rules) of the \textsf{AdaptationManager} depending on the availability of the effectors could be specified 
(\eg, \textsf{AdaptationManagerAllEffectors}, \textsf{AdaptationManagerNoRingtone} and \textsf{AdaptationManagerNoVibration}, which are
omitted here due to space restrictions). This results in different adaptation spaces, each defined by a state machine. Such  state machines define how the \textsf{AdaptationManager} reacts to context switches by employing only those adaptation rules that use available and not failed~effectors.

In the following, we use the different contextual and adaptation spaces as a justification for having multiple \textsf{ContextManager}s and  \textsf{AdaptationManager}s, respectively, to show how controllers can be adapted according to the current needs of the target system (\eg, availability of sensors and effectors).

\section{Approach} 
\label{sec:aprroach}

%\ap{Restructure this section to include details of the meta-controller. Suggestion (note that Reuse is not on focus):  3. Approach; 3.A Controller based on micro-controllers; 3.B Meta-controller; 3.C PhoneAdapter}

A self-adaptive software system needs to be considered in the context of its environment.
The system itself is composed of the target system and the controller~\cite{kephart2003}.
\rogerio{
The controller handles any changes affecting the target system or its environment by adapting the  system in order to satisfy its goals.
Adaptations can either be parametric or structural~\cite{andersson2009a,mcKinley2004}.
Parametric adaptations of monolithic controllers are more common~\cite{krupitzer2016} since structural adaptations may require reevaluation and redeployment of the system.
Thus, the structure of controllers, either open loop or closed loop, are usually static.

A challenge in self-adaptive software systems is the ability to decouple completely the controller from the target system because of their intrinsic intricacies and complexities.
This might be one of the reasons for hindering the reuse of controllers for self-adaptive software systems across different applications in which systems may provide different services with different levels of quality.
A promising solution 
for decoupling the controller and the target system is to introduce a two tier controller, which is essentially the key contribution of this paper. 
The lower tier, that would interact with the target system and the environment, is a structurally flexible controller that is able to adjust easily to the needs of the target system.
The higher tier, that would control the lower tier controller, is a sophisticated controller, like Rainbow~\cite{yuan2013,schmerl2014}, that is able to enforce the system goals by changing the behaviour and structure of  the lower tier controller.
}

\rogerio{
In self-adaptive systems, the loci of change is either the target system or its environment, but not the controller.
The controller is the locus of adaptation.
The main idea being promoted in this paper is that controllers do not need to be structurally static at development-time or run-time: controllers can both be the locus of change \textit{and} the locus of adaptation.
In the following, we present our key idea of  two tier controllers; 
first, we look into on how to build controller from micro-controllers (\ref{subsec:micro-controller-based Design}), and then, we present the idea of a meta-controller as a means to control an ensemble of micro-controllers (\ref{subsec:meta-controller}).
}

\subsection{\rogerio{Micro-controllers}}
%\subsection{Micro-controller-based Design}
\label{subsec:micro-controller-based Design}

%\fcf{Here I changed to focus to micro-controllers instead of microservices.}
The key idea being promoted is to replace a monolithic controller, like an implementation of the MAPE-K loop~\cite{kephart2003}, by an ensemble of \emph{{micro-controllers}}.
These micro-controllers would not be restricted to the implementation of services provided by the distinct stages of a MAPE-K loop such as monitoring and analysis.
Instead, the micro-controllers
would be associated with specific services that are associated with the individual stages of a controller~\cite{deLemos2017}.
The number of micro-controllers needed for implementing a controller would depend on the needs of a target system, and the granularity of the available micro-controllers.
For example, rather than having a micro-controller
implementing a full stage of the MAPE-K loop, a micro-controller could implement the services associated with integration testing~\cite{Silva2011}, or model checking~\cite{Sharifloo2013}.
It is the collective purpose of these micro-controllers to control the target system.
%These microservices are called \emph{micro-controllers}, and 
These micro-controllers can either be open or closed loop. 
An example of an open loop micro-controller could be a service for selecting an alternative solution based on utility theory.
On the other hand, an example of a closed loop micro-controller could be a pro-active monitoring service that would adjust the monitoring rate of several sensors depending on the observed operational state of the target system.
Therefore, in our proposed approach, a controller for a self-adaptive software system would be implemented as an ensemble of service-specific micro-controllers. 

Each micro-controller should be developed as a separate process for maximising independent deployment.
The communication between micro-controllers should be done through interfaces, using protocols like REST.
The operations should be standardised, mainly those that are relevant to controllers, like, configuration and deployment.
The choreography amongst micro-controllers should rely on the micro-controller implementing \textsf{Knowledge} since it should contain a consistent view that the controller should have of the target system and environment~\cite{Nii1986}. Thus, the state held by the controller could be maintained by the \textsf{Knowledge}, which allows the other micro-controllers to be stateless.
Although micro-controllers should be independent, the coordination between micro-controllers should follow quite closely the control flow of a MAPE-K loop.
Potential conflicts that might happen in the control flow should be dealt at single decision points. 
For example, the existence of several micro-controllers does not mean that all are able to access the target system.
Instead, a single micro-controller would be responsible for executing the adaptation plan.

\subsection{\rogerio{Meta-controller}}
\label{subsec:meta-controller}

\rogerio{If the controller, in addition of being the locus of adaptation is also the locus of change, there is the need of an additional controller, at a higher-level, that would manage and adapt the changes that occur at the controller.}
We have named this higher-level controller as \emph{meta-controller}.
Different from the traditional controllers that need to be tailored to the needs of distinct target systems, the sole purpose of the meta-controller would be to manage and adapt the micro-controllers that implement the controller~\cite{aberaldo2019}.
%\ap{[Rev.1] We can be more precise in the sentence that follows to avoid the impression that any "any systems that is a collection of services" can be easily managed/controlled. We can highlight that the MAPE-based workflow is well-known, so indeed the tailoring is simpler.}
\rogerio{The tailoring of the meta-controller to different ensembles of micro-controllers would be simpler because the `target system' of the meta-controller would be a collection of micro-controllers implementing, for example, the MAPE-K loop activities.
A good example of such controller would be Rainbow~\cite{yuan2013,schmerl2014}.}
%The tailoring of the meta-controller to different systems would be simpler since its `target system' would be a collection of micro-controllers. 
%\ap{[Rev.1] We can tune down the claim that follows. In fact, even though the intricacies of the target system would be *mostly* dealt by the micro-controllers, the meta-controller  also has knowledge of the actual target system, and changes in the target system may dictate changes in the controller (that is, changes in the target system may require changes in the controller; we have this situation in the PhoneAdapter use case)} 
\rdl{this section needs to be improved}
%All the intricacies associated with the actual target system would be dealt by the micro-controllers that implement the control.

In summary, the approach being proposed, that of synthesising controllers from micro-controllers, is beneficial because the different needs of a target system are addressed by different ensembles of  micro-controllers.
This is achieved because of the structural flexibility of the ensembles, which is managed by a meta-controller.
Realising micro-controllers as microservices, other benefits of our approach include the
independent development, deployment, operations, versioning and scaling of micro-controllers.
%, all of these inherited from using \ap{replace "microservives" with "micro-controllers"?}
% microservices at the controller level.  

%\vspace{.3cm}
%\section{Demonstration and Evaluation}
\section{Demonstration}
\label{sec:demonstration}

In this section, we \rogerio{demonstrate} our approach for synthesising controllers from micro-controllers.
More specifically, 
Section~\ref{sec:AnExample} presents how the \phoneAdapter controller was restructured based on micro-controllers.
Section~\ref{sec:phoneAdapterMicroControllers} details the implementation and shows how \phoneAdapter can be integrated into the Android Framework,
% and how we mapped code elements from the framework to work with microservices concepts.
and Section~\ref{sec:phoneAdapterMetaController} presents an overview of the \phoneAdapter meta-controller.

\subsection{PhoneAdapter Case Study}
\label{sec:AnExample}

\begin{figure}[t]
    \centering
    \includegraphics[width=0.55\columnwidth]{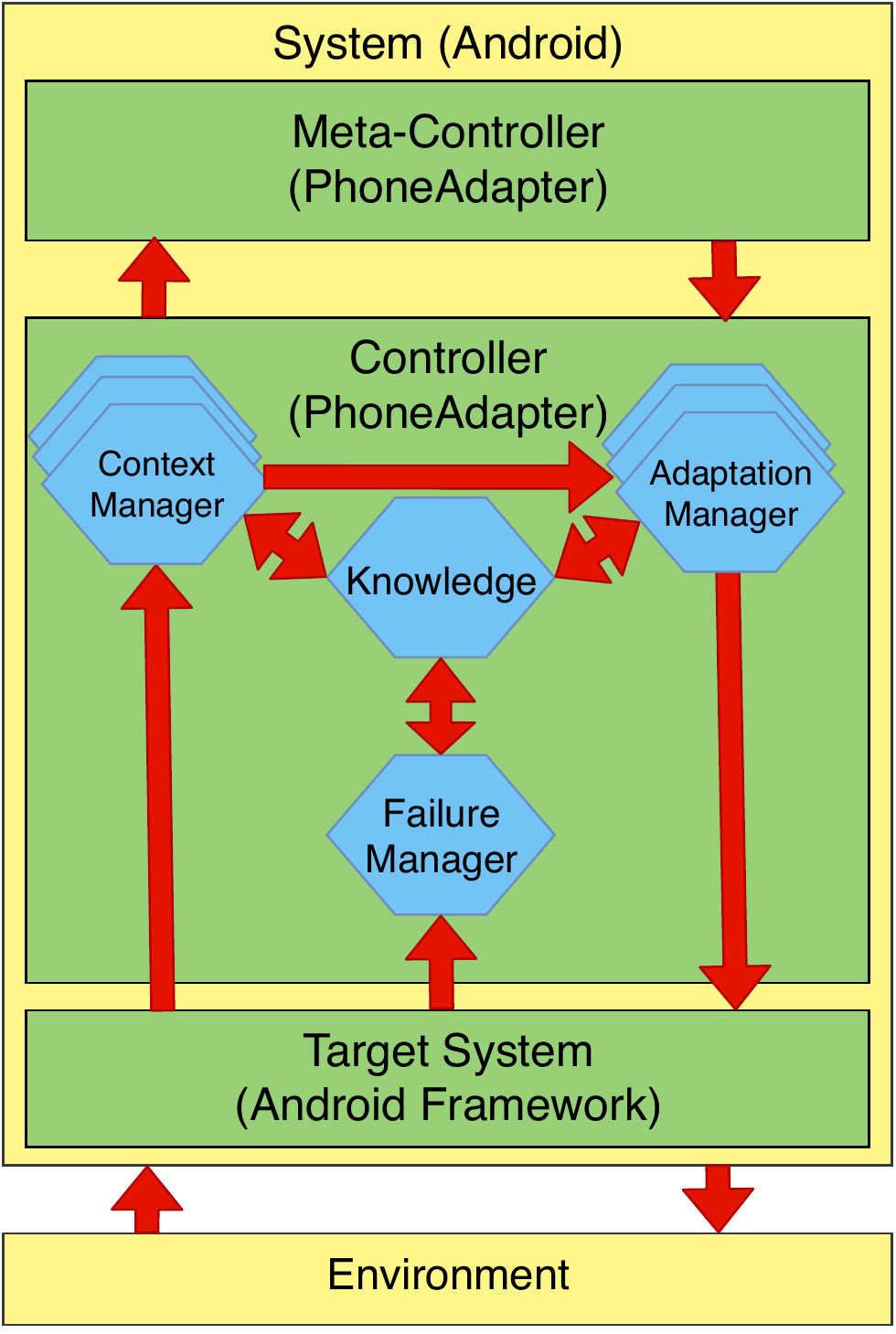}
    \vspace{-.25cm}
    \caption{PhoneAdapter implemented with micro-controllers.}
    \label{fig:PhoneAdapter_MultipleMicroC}
    \vspace{-1em}
\end{figure}

% \fcf{Highlight that PhoneAdapter control services of the target system (e.g. BT, GPS, audio volume, vibration etc.}
The controller for the \emph{PhoneAdapter} case study consists of two components, which incorporate most of the functionality of a MAPE-K loop (see Figure~\ref{fig:phoneAdapterArchitecture}). 
The refactoring of that controller for a solution based on micro-controllers,
%\ap{Double-check if \newPhoneAdapter is a good name. If agreed, revise all occurrences from hereafter \note{the command is defined in commands.tex}}
which we hereafter call \newPhoneAdapter, is shown in Figure~\ref{fig:PhoneAdapter_MultipleMicroC}.
Instead of a single pair of \textsf{ContextManager} and \textsf{AdaptationManager}, the design of \newPhoneAdapter consists on the combination of several micro-controllers that are able to capture variants of the original \textsf{ContextManager} and \textsf{AdaptationManager}. These variants depend on sensor and effectors failures. 
The motivation for having several variants of managers is to save device battery by avoiding to unnecessarily activate \rogerio{failed} sensors, and to avoid to adapt the target system without success due to failed effectors.

In addition to a combination of \textsf{ContextManager} and \textsf{AdaptationManager} variants, the \newPhoneAdapter controller includes the \textsf{Knowledge} and  \textsf{FailureManager} micro-controllers.
\rdl{the following sentence could be removed}The first one is used for storing the models of the target system and system environment, and the latter is used to monitor the operational status of the sensors and effectors.
These two components could be generic micro-controllers that could be reused across several Android applications that use sensors and effectors provided by the Android Framework. 
By externalising the knowledge to a dedicated micro-controller, we consider the other micro-controllers to be stateless.

\rogerio{The role of the meta-controller would be to manage the controller in terms \bento{of} (re)configuring  the micro-controllers depending on the operational status of the sensors and effectors.}
%The role of the meta-controller would be to manage the controller in terms of (re)composing the micro-controllers and (re)configuring each individual micro-controller depending on the operational status of the sensors and actuators.
\rogerio{This consists of introspecting the \textsf{Controller}'s \textsf{Knowledge}, evaluating the current configuration of the micro-controllers deployed in the \textsf{Controller}, deciding whether the \textsf{Controller} needs to be adapted, selecting the most appropriate micro-controllers from a repertoire, and finally, reconfiguring the micro-controllers without affecting the operation of the target system.}
%This consists of introspecting the \textsf{Controller}'s \textsf{Knowledge}, evaluating the current composition and configurations of the micro-controllers deployed in the \textsf{Controller}, deciding whether the \textsf{Controller} needs to be adapted, selecting the most appropriate micro-controllers from a repertoire, and finally, recomposing and reconfiguring the micro-controllers without affecting the operation of the target system. 

For the implementation of the case study, we used the original source code\footnote{Available at \scriptsize{\url{https://github.com/californi/PhoneAdapter_Original}}} of the \phoneAdapter provided by~\citet{liu2013}, which comprises 5,642 lines of code (1,345 referring to adaptation) and 21 classes (8 referring to adaptation).
Moreover, we have used Android Studio\footnote{\url{https://developer.android.com/} -- accessed on May, 2020.} IDE, a physical mobile device,\footnote{Samsung-gt\_p5200-3200118b11f87000 (API level 8)}
and a virtual device\footnote{CPU/ABI: 
Google Play Intel Atom (x86) (API level 29)} to deploy the software artefacts.
The \newPhoneAdapter\footnote{Available at \scriptsize{\url{https://github.com/californi/PhoneAdapter_MicroControllersSA}}} 
comprises
9,821 lines of code (5,483 referring to adaptation) and 45 classes (32 referring to adaptation).
%\ap{emphasise here the refactoring of the controller as the reason for larger code size.}\bento{OK}
\fabiano{Approximately, 3,714 LOCs and 20 classes refer to added variants and duplicate code present in them.}

% \ap{Restructure this section. We need to make clear how adaptation rules for the synthesised controller are processed.}

% \ap{We may insert table with adaptation rules for the controller} \bento{ok}

%\ap{Firstly, we may present the list of micro-controller and operations for the PhoneAdapter micro-controllers cf. Table~\ref{tabMicroserviceOperations}, and the operations for the  meta-controllers cf. Table~\ref{tabMetaControllerOperations}; then, we can explain how deployment happens.}

\subsection{Implementation and Deployment of Micro-controllers} \label{sec:phoneAdapterMicroControllers}

%\fabiano{As previously mentioned in this paper, we demonstrate our approach in a Android application.}
The Android Framework (cf.~Figure~\ref{fig:PhoneAdapter_MultipleMicroC}) provides several Application Programming Interfaces (API) on top of which mobile applications are developed.
In general, the application framework layer and other underlying layers (\eg, the Android Runtime and Hardware Abstraction layer) come pre-installed on Android devices. 
These layers provide abstractions that developers use to implement their applications. 
% to concentrate only in more abstract layers. 
% Thus, when a developer needs to construct a new application, she only uses libraries in a higher level. 
To test and execute an application, either a physical or virtual Android device is used.
% An Android device could be as a physics device as virtual device.
With respect to Android development, which is mostly based on the Java language, developers are able to extend any available class from the framework using different types of features (\eg, dealing with service communication). %\tv{which components? Or do we ``just'' develop and install apps?}
%
%When developing Android applications, 
The usage of activities, intents, services, and content providers is intrinsic for any application.
%\tv{Do we need these Android details to describe our approach?}
%\brs{We agree!   but maybe we need to include some information in a place (e.g. android services, message intent )}
%\fcf{these concepts may be described in a Background section (to be written).}
%\tv{We should keep those aspects that are relevant for the mapping of microservices to Android concepts and that are mentioned later on in this section.}

%\vspace{.3cm}
%\noindent\textbf{Activity.} 
%An activity is a graphical user interface with a single screen and it defines a simple life cycle for handling interruptions. An application can define one or more activities to handle different phases of the program.
%\vspace{.3cm}
%
%\noindent\textbf{Intent.} 
%An intent is a mechanism for describing a specific action, such as ``pick a photo'', or ``send a message''. In Android, \bento{any interaction} %\tv{Interactions?} 
%goes through intents that may invokes activities or handle the communication between components inside the system.
%\vspace{.3cm}
%
%\noindent\textbf{Service.} 
In \rogerio{the Android Framework}, a service is a task that runs without any direct user interaction. %done! \tv{rephrase:} 
It can also be developed by using different features (\eg, BroadcastReceiver, IntentService) and types (\ie, foreground, background and bind-Service), \eg, by using background services or tailored services\rdl{this can be removed and the paragraphs combined}. 
%\tv{What are these features and types?}
%
%\vspace{.3cm}
% \noindent\textbf{Content Provider.} 
%A content provider is a set of APIs to write and read data. This is used to share global data between applications.
%\vspace{.3cm}
%\tv{The deployment of microservices should come later when discussing how microservice concepts have been mapped into Android. We shoud mention this mapping explicitly.}
%\brs{We could put ``deployment of microcontroller'', instead of only ``deployment''}

\begin{figure}[!ht]
    \centering
    \includegraphics[width=.9\columnwidth]{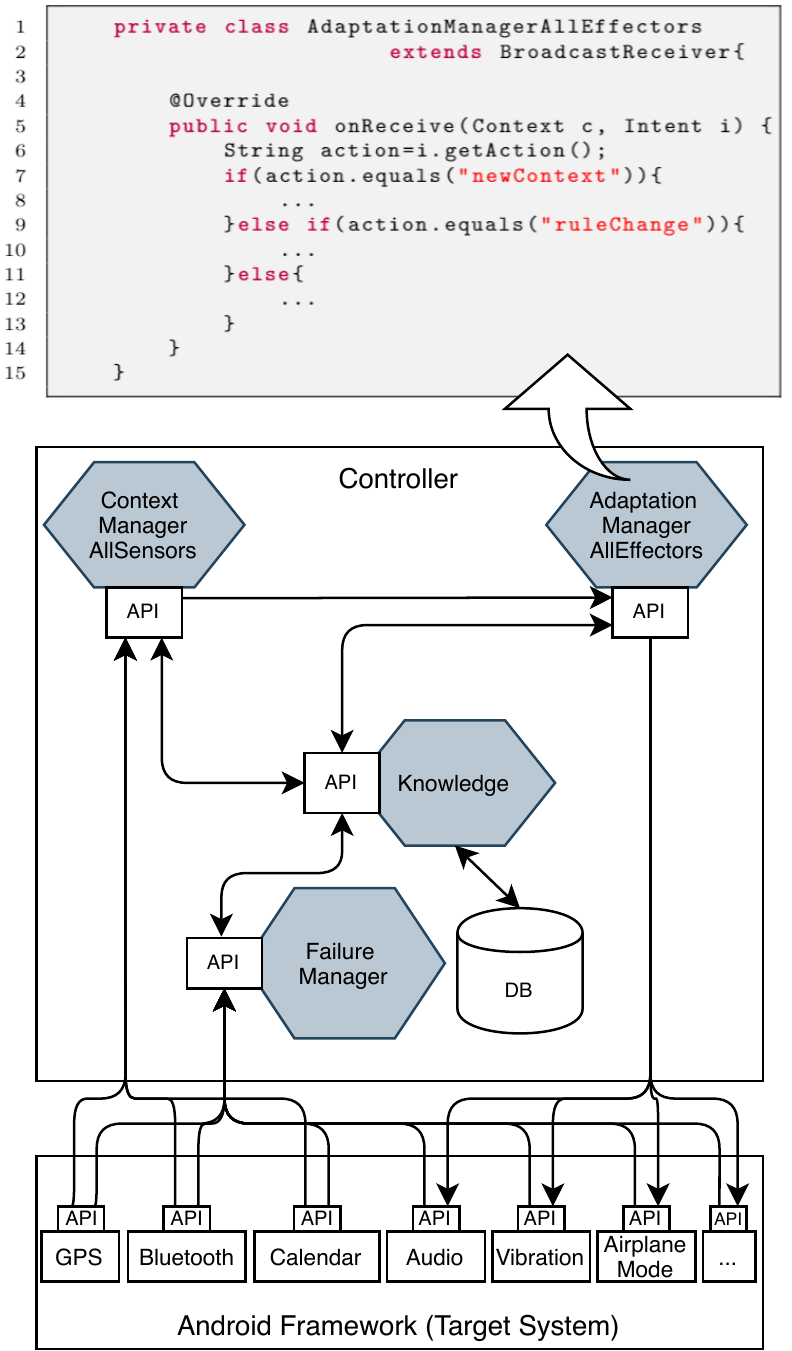}
    %\vspace{-.3cm}
    \caption{A possible configuration of the \newPhoneAdapter. % controller.
    %\note{Link to figure of FSMs and rules: \url{https://drive.google.com/file/d/1uc_-hP7hDund5Ul4DIb67FijAi0RbS2W/view?usp=sharing}}
    }
    \label{fig:ControllerConfiguration}
    \vspace{-1em}
\end{figure}

%\ap{Revise the following description to explain how PhoneAdapter is currently structured (with BroadcastReceiver and IntentService)}\bento{ok}
In the \phoneAdapter synthesised with micro-controllers (\ie, \newPhoneAdapter), each micro-controller is implemented as class that extends the \texttt{BroadcastReceiver} class, and also as class that extends \texttt{IntentService}\footnote{\texttt{BroadcastReceiver} and \texttt{IntentService} are Resources to define microservices in the Android framework. 
By using the same resources, it is possible to define REST API available in servers outside Android framework.}. % from the Android Framework. 
At the top of Figure~\ref{fig:ControllerConfiguration}, we show a code snippet of a micro-controller implemented as a \texttt{BroadcastReceiver} class.
In line 5, the \texttt{onReceive} event is responsible for handling all messages that are generated by other services (\eg, messages sent by one of the various the \textsf{ContextManager} components). 
Finally, in lines 7 and 9 the conditions are used to handle messages captured by the \texttt{onReceive} event. 
\bento{Both classes---\texttt{BroadcastReceiver} and \texttt{IntentService}---have the same goal, but have their differences related to the life cycle of services and the Android Environment. For example, \texttt{BroadcastReceiver} works as a listener of messages sent and received from the Android Environment, whereas  \texttt{IntentService} works as a service that is invoked by \eg, other services or activities from graphical user interfaces.} \rdl{for the sake of space, this additional text from Bento could be removed}

The variations of \newPhoneAdapter's \textsf{ContextManager} and \textsf{AdaptationManager} micro-controllers were developed using tailored services similar to the one presented in the code snippet of Figure~\ref{fig:ControllerConfiguration}. 
Table~\ref{tabMicroserviceOperations} presents the list of micro-controllers and their operations. 
The inputs and output for each operation are also shown in the table.
Operations defined by each micro-controller can be mapped to internal code elements. 
For example, considering the \textsf{AdaptationManagerAllEffectors} micro-controller (see code snippet in Figure~\ref{fig:ControllerConfiguration}),
its \texttt{/processNewContext} and \texttt{/processRuleChange} operations are mapped to the \texttt{onReceive} method that represents an event in Android \rogerio{Framework}.
Given that a single method (\ie, \texttt{onReceive}) is mapped to two micro-controller operations, internal decisions (based on action values embedded in the \texttt{Intent} object) will define the correct logic, that associated with a particular operation, to be processed.

The deployment of the micro-controllers 
is realised through the two following operations: 
\texttt{unregisterReceiver(}...\texttt{)} and
\texttt{registerReceiver(}...\texttt{)}.
The former deactivates the current micro-controller (\eg, \textsf{ContextManagerAllSensors}), and the latter activates an alternative micro-controller 
that is in conformance with the current context (\eg, \textsf{ContextManagerNoGPS} when the GPS sensor is malfunctioning).

As already mentioned, assuming that sensors and effectors may fail, we could have alternative \textsf{ContextManagers} and \textsf{AdaptationManagers} composing the controller. 
The bottom part of Figure~\ref{fig:ControllerConfiguration} represents the \newPhoneAdapter's micro-controllers and their data flow (the arrows follow the same flow we presented in Figure~\ref{fig:PhoneAdapter_MultipleMicroC}).
If failures happen \eg,
in the GPS sensor and in the Ringtone effector, the controller will be reconfigured with two alternative micro-controllers (namely, \textsf{ContextManagerNoGPS} and \textsf{AdaptationManagerNoRingtone}).
%The replacement of the \textsf{ContextManager} occurs when the \textsf{FailureManager} detects and signals a sensor failure (by using the  \textsf{/verifySensors} operation cf. Table~\ref{tabMicroserviceOperations})
The replacement of the \textsf{ContextManager} occurs when the \textsf{FailureManager} micro-controller detects and stores a sensor failure into the \textsf{Knowledge}
(by using the  \textsf{/verifySensors} operation, cf. Table~\ref{tabMicroserviceOperations}). Such a failure is retrieved by the \textsf{MetaController} (see Section~\ref{sec:phoneAdapterMetaController} for details) by using the  \textsf{/sensorsFailure} operation (cf. Table~\ref{tabMetaControllerOperations}); 
the \textsf{MetaController} then analyses the failure and adapts the Controller accordingly (\eg, by activating a new \textsf{ContextManager}).
The \textsf{FailureManager} and the \textsf{MetaController} handle the \textsf{AdaptationManager} variants in a similar manner by using the \textsf{/verifyEffectors} and \textsf{/effectorsFailure} operations, respectively.

\begin{table*}[!ht]
	\caption{Operations of the micro-controllers in \newPhoneAdapter.}
%	\todo{Revise this table... remove Meta-controller... keep one instance of each microcontroller and list its operations}
			\centering
			%\small

            %\vspace{-.2cm}

	        \setlength{\tabcolsep}{1pt}
            %\scriptsize   
            % \small
            
            \fontencoding{T1}
            %\fontfamily{\ttdefault}
            \fontfamily{\rmdefault}
            \fontseries{m}
            \fontshape{n}
            \fontsize{7.7}{6.7}
            \selectfont   
            
            %\vspace{-0.2cm}

			\label{tabMicroserviceOperations}
\begin{tabular}{llp{5.5cm}p{6cm}} \toprule
\rowcolor{lightgray!50}
\textbf{Micro-controller}                  & \textbf{Operation} & \textbf{Input}                                  & \textbf{Output}                                      \\ \midrule

ContextManagerAllSensors               & /generateContext   & -                                               & An intent message with: GPS, Bluetooth, Calendar data \\

\rowcolor{lightgray!50}
               & /sensingBluetooth  & -                                               & A bluetooth collection                               \\

               & /locationListener  & -                                               & Latitude, Longitude data                             \\

\rowcolor{lightgray!50}
ContextManagerNoGPS                    & /generateContext   & -                                               & An intent message with: Bluetooth, Calendar data      \\

                    & /sensingBluetooth  & -                                               & A bluetooth collection                               \\

\rowcolor{lightgray!50}
ContextManagerNoBluetooth              & /generateContext   & -                                               & An intent message with: GPS, Calendar data            \\

              & /locationListener  & -                                               & Latitude, Longitude data                             \\

\rowcolor{lightgray!50}
ContextManagerNoGPSNoBluetooth         & /generateContext   & -                                               & An intent message with: Calendar data                 \\

AdaptationManagerAllEffectors          & /processNewContext & newContext: GPS, Bluetooth and Calendar data    & New state in volume; vibration; airplane             \\

\rowcolor{lightgray!50}
          & /processRuleChange        & ruleChange                                      & Back to general state                                \\

AdaptationManagerNoRingtone            & /processNewContext & newContext: GPS and Bluetooth and Calendar data & New state in vibration; airplane                     \\

\rowcolor{lightgray!50}
            & /processRuleChange  & ruleChange                                      & Back to general state                                \\

AdaptationManagerNoVibration           & /processNewContext & newContext: GPS and Bluetooth and Calendar data & New state in volume; airplane                        \\

\rowcolor{lightgray!50}
           & /processRuleChange & ruleChange                                      & Back to general state                                \\
%AdaptationManagerNoRingtoneNoVibration & /processNewContext & newContext: GPS and Bluetooth and Calendar data & New state in airplane                                \\
%AdaptationManagerNoRingtoneNoVibration & /ruleChange        & ruleChange                                      & Back to general state                                \\

Knowledge           & /NewSensorContext & SensorContext: GPS status and Bluetooth status & An intent message with a new generated sensors context \\

\rowcolor{lightgray!50}
           & /NewEffectorData & EffectorData: Ringtone status, vibration status, volume number & An intent message with a new generated effectors data \\

FailureManager                         & /verifySensors     & GPS, Bluetooth status                           & An intent message localising a sensor failure                                \\

\rowcolor{lightgray!50}
                         & /verifyEffectors   & Audio, Vibration                                & An intent message localising an effector failure                              \\ \bottomrule
%FailureManager                         & /locationListener  & -                                               & GPS status                                           \\
%FailureManager                         & /bluetoothStatus   & -                                               & Bluetooth status                                     \\
%FailureManager                         & /effectorsStatus   & -                                               & Audio, Vibration               
\end{tabular}
\end{table*}

\vspace{0.2cm}
\subsection{The \newPhoneAdapter Meta-controller}  \label{sec:phoneAdapterMetaController}

%\ap{Create a subsection for meta-controller (analogous to Approach section).}

%\fcf{[TODO] Double-check how to address the meta-controller here. Do we have details of its implementation in PhoneAdapter that could be inserted here?} \bento{ok}

% \fcf{Bento is creating a table of adaptation rules for the controller... it includes rules for replacing the ContextManager, and rules for replacing the AdaptationManager}\bento{ok}

In Section~\ref{sec:aprroach}, we argued that a meta-controller could be developed to manage the architectural configuration of the controller depending, for example, on the operational status of the sensors and effectors. 
In our demonstration, the meta-controller reconfigures the controller according to the \newPhoneAdapter's needs.
\rogerio{In the context the example mentioned above}, the \textsf{MetaController} deactivates  \textsf{ContextManagerAllSensors} and activates  \textsf{ContextManagerNoGPS}; similarly, it deactivates  \textsf{AdaptationManagerAllEffectors} and activates  \textsf{AdaptationManagerNoRingtone}, considering that these micro-controllers are stateless.

Tables~\ref{tabAdaptationRulesManagingSystemContextManagers} and \ref{tabAdaptationRulesManagingSystemAdaptationManagers} show the adaptation rules processed by the \textsf{MetaController}, regarding the \textsf{ContextManager} and the \textsf{AdaptationManager} micro-controllers designed for \newPhoneAdapter, respectively. 
For example, the \emph{ActivateNoBlutooth} rule (rule (c) in Table~\ref{tabAdaptationRulesManagingSystemContextManagers}) establishes that  \textsf{ContextManagerNoBluetooth} must be activated when the GPS sensor is enabled and the Bluetooth sensor is not (cf. the predicate defined in column ``Full Predicate'').

\begin{table*}[!ht]
    \centering
    \caption{Adaptation rules for the \textsf{MetaController} related to \textsf{ContextManager}.}

            \vspace{-.2cm}

	        \setlength{\tabcolsep}{1pt}
            %\scriptsize   
            % \small
            
            \fontencoding{T1}
            %\fontfamily{\ttdefault}
            \fontfamily{\rmdefault}
            \fontseries{m}
            \fontshape{n}
            \fontsize{7.7}{6.7}
            \selectfont   
            
            %\vspace{-0.2cm}
    \label{tabAdaptationRulesManagingSystemContextManagers}

\begin{tabular}{cp{3.2cm}p{2.5cm}p{2.7cm}p{5cm}l}
\rowcolor{lightgray!50}
    \toprule
    \textbf{Id} & \textbf{Rule Name} & \textbf{Current ContextManager} & \textbf{New ContextManager} & \textbf{Full Predicate} & \textbf{Output (Controller Configuration)} \\ 
    \midrule
    a & Activate AllSensors & Any & AllSensors  & GPS.isEnabled() \&\& Bluetooth.isEnabled() & AllSensors is enabled \\
\rowcolor{lightgray!50}    
    b & Activate NoGPS & Any &  NoGPS &  !GPS.isEnabled() \&\& Bluetooth.isEnabled() & NoGPS is enabled \\
    
    c & Activate NoBluetooth & Any & NoBluetooth &  GPS.isEnabled() \&\& !Bluetooth.isEnabled() & NoBluetooth is enabled \\
\rowcolor{lightgray!50} 
    d & Activate NoBluetoothNoGPS & Any & NoBluetoothNoGPS &  !GPS.isEnabled() \&\& !Bluetooth.isEnabled() &  NoGPSNoBluetooth is enabled \\ \bottomrule

\end{tabular}
\end{table*}    
    
\begin{table*}[!ht]
    \caption{Adaptation rules for the \textsf{MetaController} related to \textsf{AdaptationManager}.}

			\centering
			%\small

            \vspace{-.2cm}

	        \setlength{\tabcolsep}{1pt}
            %\scriptsize   
            % \small
            
            \fontencoding{T1}
            %\fontfamily{\ttdefault}
            \fontfamily{\rmdefault}
            \fontseries{m}
            \fontshape{n}
            \fontsize{7.7}{6.7}
            \selectfont   
            
            %\vspace{-0.2cm}
    \label{tabAdaptationRulesManagingSystemAdaptationManagers}

\begin{tabular}{cp{3.8cm}p{2.1cm}p{3cm}p{4.5cm}l}
\rowcolor{lightgray!50}
\toprule
     \textbf{Id} & \textbf{Rule Name} & \textbf{Current AdaptationManager} & \textbf{New AdaptationManager} & \textbf{Full Predicate} & \textbf{Output (Controller Configuration)} \\  
    \midrule
    e & Activate AllEffectors & Any & AllEffectors &  Vibration.IsON \&\& Audio.Volume \textgreater 0 & AllEffectors is enabled \\
\rowcolor{lightgray!50}
    f & Activate NoRingtone & Any & NoRingtone &  Vibration.IsON \&\& !Audio.Volume \textgreater 0  & NoRingtone is enabled \\
    
    g & Activate NoVibration & Any & ManageNoVibration &  Vibration.IsOFF \&\& Audio.Volume \textgreater 0  & NoVibration is enabled \\
\rowcolor{lightgray!50}    
    h & Activate NoRingtone NoVibration & Any & NoRingtoneNoVibration &  Vibration.IsOFF \&\& !Audio.Volume \textgreater 0 & NoRingtoneNoVibration is enabled \\ \bottomrule
    
\end{tabular}
\end{table*}

\begin{lstlisting}[basicstyle= \ttfamily\footnotesize, caption={\textit{Meta-Controller
component}}: Enabling and disabling variants of \textsf{ContextManager} and \textsf{AdaptationManager} (code only partially shown).,label={lst:metaController}]
public class MetaControllerBroadcastReceiver 
                        extends BroadcastReceiver {
    @Override
    public void onReceive(Context c, Intent i) {
        if (i.action.equals("sensorsFailure")) {
            activatingContextManager(c, i);
        } 
        else 
          if (i.action.equals("effectorsFailure")){
            activatingAdaptationManager(c, i);
        } else {
            ...
        }
    }
    
    private void activatingContextManager(c, i) {
        if (i.action.equals("AllSensors")){
            ...
        } else if (i.action.equals("NoGPS")){
            ...
        } else {
            ...
        }
    }
    
    private void activatingAdaptationManager(c, i){
        if (i.action.equals("AllEffectors")) {
            ...
        } else if (i.action.equals("NoRingtone")){
            ...
        } else {
            ...
        }
    }
}
\end{lstlisting}

Similarly to the micro-controllers described in the previous section, 
the \textsf{MetaController} in the \newPhoneAdapter application is characterised as an Android \rogerio{Framework} microservice,
%and its API
as shown in Table~\ref{tabMetaControllerOperations}.
\rogerio{The activation/deactivation of micro-controllers is managed by the \textsf{MetaController} service, with support of the \textsf{FailureManager} service, depending whether the sensors and effectors are working or not.}
\rogerio{For example, the code snippet
%\footnote{Due to space limitations, the code is only partially shown.} 
of Listing~\ref{lst:metaController} shows how the \textsf{MetaController} reacts to failures 
signalled by the \textsf{FailureManager}.}
%, and how these affect the configuration of \textsf{ContextManager} and \textsf{AdaptationManager}.
First, \rogerio{in lines 5 to 13 it identifies the type of failure that occurred.} 
For example, if the GPS sensor stops working, the \textsf{MetaController} enables \textsf{ContextManagerNoGPS} (Figure~\ref{fig:ContextManagerNoGPS}) and disables \textsf{ContextManagerAllSensors} (line 19). 
The same occurs in the event of effectors failures
%, \eg, if all effectors are working then the \textit{AdaptationManagerAllEffectors} is enabled and so on 
(lines 27-33).

\begin{table*}[!ht]
	\caption{Operations of \textsf{MetaController} in \newPhoneAdapter.
	%\fcf{[TODO] Check where this table should be mentioned in the text.} \bento{I put with the explanation of the meta-controller code snippet.}
	}
			\centering
			%\small

            %\vspace{-.2cm}

	        \setlength{\tabcolsep}{2pt}
            %\scriptsize   
            % \small
            
            \fontencoding{T1}
            %\fontfamily{\ttdefault}
            \fontfamily{\rmdefault}
            \fontseries{m}
            \fontshape{n}
            \fontsize{7.7}{6.7}
            \selectfont   
            
            \vspace{-0.2cm}

			\label{tabMetaControllerOperations}
\begin{tabular}{llp{7cm}p{5cm}} \toprule
\rowcolor{lightgray!50}
\textbf{Meta-controller}                  & \textbf{Operation} & \textbf{Input}                                  & \textbf{Output}                                      \\ \midrule

MetaController                        & /sensorsFailure    & failure message regarding GPS or Bluetooth status                           & A new ContextManager                                 \\

\rowcolor{lightgray!50}
                        & /effectorsFailure  & failure message regarding Audio or Vibration                                & A new AdaptationManager                              \\
\bottomrule

\end{tabular}
\end{table*}

\section{Evaluation and Discussion} \label{sec:evaluation}

%{\LARGE \red{[Fabiano] Writing a new version of this section... the original is saved at evaluationOLD.tex}}

%\fcf{[Rev.1: "not relevant evaluation. It is performed in a technological stack that is not adequate to evaluate the microservices proposal"] 
    %We can evolve the evaluation in terms of quantitative and qualitative viewpoints.    
    %Regarding the platform and technological limitation, Android may not be adequate. Two possible actions: (1) reduce the claim for microservices, and generalise towards flexible controllers based on microcontrollers (microservices is an option for implementing them); or (2) find another target system that allows for a proper modularisation with microservices.}
%\fcf{See the evaluation plan in meetingNotex.tex}

This section presents 
a qualitative evaluation of both \textit{PhoneAdapter} versions (Section~\ref{sec:qualitativeEvaluation}), 
the threats to the validity of our implementation and its initial evaluation (Section~\ref{sec:threats}), and
brings some further discussions regarding the reusability of micro-controllers (Section~\ref{sec:furtherDiscussion}).

\subsection{Qualitative Evaluation} \label{sec:qualitativeEvaluation}

%\ap{Write introductory paragraph to say that we will focus our discussion on the structural flexibility, and then discuss how it brings benefits in terms of: (i) handling changes at run-time (adaptability); (ii) handling changes at design time (maintainability) }

\fabiano{
This section presents an initial, qualitative evaluation of our approach based on the \phoneAdapter case study. 
We focus on \rogerio{the structural flexibility} of the controller and how it supports 
(i) handling changes at run-time, % (adaptability)
and 
(ii) handling changes at design time. % (maintainability)
We provide some examples from the two implementations (the \rogerio{\phoneAdapter}~\cite{liu2013}  and \newPhoneAdapter) to justify our arguments.
}

% \ap{Revise / rephrase to focus on flexibility (and avoid focus on self-healing)}

\vspace{.2cm}

%\noindent\textbf{Handling changes at run-time:} A controller that can be adapted to properly handle scenarios such as the one illustrated in  Figures~\ref{fig:ContextManagerAllSensors} and \ref{fig:ContextManagerNoGPS}---and others discussed throughout this paper---is inherently flexible. This is observed on our proposed solution based on micro-controllers. For example, when a known failure occurs (\eg, in a sensor, in an effector, or in both), the \newPhoneAdapter is able to reconfigure at run-time and handle it, whereas this would bring some implications to the original \phoneAdapter. This is discussed next.

\rogerio{\noindent\textbf{Handling changes at run-time:} 
A controller is structurally flexible at run-time, if it is able to adapt its structure in order to handle the context and failure scenarios as the ones illustrated in  Figures~\ref{fig:ContextManagerAllSensors} and \ref{fig:ContextManagerNoGPS}.
We have shown this to be possible using a solution based on micro-controllers.
For example, \newPhoneAdapter is able to reconfigure at run-time in the presence of failures (\eg, in a sensor, in an effector, or in both), whereas such a failures in the \phoneAdapter would generate some undesirable behaviours.
}

%\vspace{-0.1cm}

\begin{list}{\labelitemi}{\leftmargin=1em}

    \item \textbf{Original \phoneAdapter: }
    Given a failure in the GPS sensor, \phoneAdapter could generate inconsistent states, \eg, a Latitude of 0.0; and a Longitude of 0.0. Such a failure also impacts on the constantly monitoring of GPS data performed by the \textsf{ContextManager}. 
    In addition, even with the occurrence of GPS failures, all 
    contextual 
    rules will always be considered and processed by the \textsf{AdaptationManager}. 
    Overall, consequences could be: 
    inconsistent sensor data that would be generated (\eg, wrong location); 
    unnecessary processing, and hence battery usage (\eg, continually trying to access GPS sensor); 
    and setting of inconsistent states (\eg, in ``\textsf{Driving}'' state is set even when ``\textsf{Driving Fast}'' holds, \ie, speed is wrongly computed).
    
    \item \textbf{\newPhoneAdapter: } 
    If the same failure in the GPS sensor occurs, the  \textsf{FailureManager} micro-controller would identify this specific failure, and the \textsf{MetaController} would mitigate its consequences. 
    To do so, the \textsf{FailureManager} would signal the failure to the \textsf{Knowledge} service, 
    and \textsf{MetaController} would detect that
    change and replace 
    the \textsf{ContextManagerAllSensors} with
    the \textsf{ContextManagerNoGPS}. 
    Consequently, the monitoring would stop generating GPS data and no longer the \textsf{AdaptionManager} would retrieve inconsistent data. 
    In addition, while the \textsf{ContextManagerNoGPS} is active, the contextual 
    rules regarding GPS would not be used.  
    This scenario is illustrated in Figures~\ref{fig:ContextManagerAllSensors} and \ref{fig:ContextManagerNoGPS}. 
    Note that in the \textsf{ContextManagerNoGPS} FSM the number of transitions is smaller when compared to its predecessor (\ie, \textsf{ContextManagerAllSensors}).

\end{list}

%\noindent\textbf{Original version.} ...

%\vspace{0.3cm}

% \noindent\textbf{New version.} ...

%\noindent\textbf{Handling changes at design time:} This is illustrated in terms of the original and \newPhoneAdapter implementations as follows.

% ALTERNATIVE TEXT: 
\rogerio{
\noindent\textbf{Handling changes at design time:}
A controller is structurally flexible at design time, if a controller can be easily replaced by an alternative one.
If a controller is complex in a way that incorporates combinatorial behaviours involving context changes and failures, 
%like the managers of \phoneAdapter, 
then this may hinder their replacement. 
On the other hand, micro-controllers that are able to express simple behaviours can be easily replaced.
%For example, \newPhoneAdapter allows to replace the micro-controller \textsf{ContextManagerNoGPS} because of its functional simplicity.
}

%Only one ContextManager and/or AdaptationManager is possible to deploy each time. It is not possible to change at runtime.	
    
%Specific Changes in code-snippets of AdaptationManager; ContextManager	
    
%Monolith solution, several changes but easy to keep.

\begin{list}{\labelitemi}{\leftmargin=1em}

    \item \textbf{Original \phoneAdapter: }
    To deal with failures in sensors and/or effectors, or to process differently context values or contextual  
    rules, in the original version it would be necessary to tailor the \textsf{ContextManager} and/or the \textsf{AdaptationManager} components to these changes or new requirements.
    \rogerio{This makes these components more complex, thus hindering their replacement.}
    %\rdl{the following text doesn't fit with the above argument: This would allow to concentrate the changes in a small set of classes.     On the other hand, programming errors could cause a malfunction of the entire \textsf{ContextManager} or \textsf{AdaptationManager}, \fabiano{since single (or small sets of) modules embed all the behaviour and its possible variations.}}

    \item \textbf{\newPhoneAdapter: }
    \fabiano{
    If the specified behaviour changes, or new requirements raise, the developer needs to include other types of micro-controllers 
    (\eg, alternative \textsf{ContextManager} or \textsf{AdaptationManager}), 
    The micro-controllers are independently developed and their coordination is localised in the \textsf{MetaController}. 
    Changes can also be performed in a localised manner in particular variants of the micro-controllers.
    }

\end{list}

%\noindent\textbf{Original Version.} ...

%\vspace{0.3cm}

%\noindent\textbf{New Version.} ...

%\noindent\textbf{Quantitative evaluation - Overhead:}
%Original Version	
%    Given the scenario: General -> Office -> Meeting, X operations are performed.	
%    AllSensors	
%    Failure in GPS: AllSensors
%New Version	
%    Given the scenario: General -> Office -> Meeting, Y operations are performed.	
%    AllSensors	
%    Failure in GPS: No GPS

%\ap{Remove the Trade-offs label and shorten the paragraph to summarise what has been discussed}

\fabiano{
In summary, in the above discussion we 
compared both versions of \phoneAdapter in order to identify the benefits our approach provides in terms of the structural flexibility of the devised controllers.
Our approach enables the definition of new micro-controllers to compose controllers according to new requirements, without requiring substantial code changes. 
Furthermore, micro-controllers and their variants can be activated and deactivated at run-time in order to enable the controller to properly operate on its target system.

%\ap{Rephrase because we added behaviour not present in the original version}
On the downside, 
it is a well-known issue that higher maintainability implies higher performance overhead~\citep{parnas1972}. 
For example, by applying our approach to restructure the original \phoneAdapter controller as set of independent and coordinated micro-controllers (in the \newPhoneAdapter), extra processing is necessary due to the presence of at least one additional--- and required ---control loop: the one of the \textsf{MetaController} that continually monitors and eventually adapts the controller.
On the other hand, note that the original \phoneAdapter only operates with all sensors (cf. Figure~\ref{fig:ContextManagerAllSensors}).
All in all, even with the presence of a meta-controller and \eg, the added \textsf{FailureManager} micro-controller in the new version, depending on the \textsf{ContextManager} that is active (\eg, \textsf{ContextManagerNoGPSNoBluetooth}), that version could have a low---possibly no---overhead when compared to the original implementation.}

\vspace{-0.1cm}

\subsection{Threats to Validity}\label{sec:threats}

In this paper we present a novel approach, based on \fabiano{micro-controllers}, for the design and deployment of controllers for self-adaptive software systems. 
The feasibility of the approach, 
\fabiano{with focus on flexibility of controllers},
has been demonstrated through a case study based on an Android application.
However, there are some assumptions that may affect the validity of the results. 

Regarding internal validity, Android is not the most appropriate environment to implement faithfully \fabiano{micro-controllers as microservices}.
In spite of this, in our implementation of  \phoneAdapter (\ie, \newPhoneAdapter), we have used all the features available on Android to incorporate the basic principles associated with microservices.
Another internal validity threat is related to the claims regarding reuse.
Evidence for supporting this claim should have been obtained by using a micro-controller in different architectural configurations, across various target systems from different application domains.
However, in our experiments, we have just shown that micro-controllers can be used across different architectural configurations of a controller for one target system  
\fabiano{(we further discuss reusability issues in Section~\ref{sec:furtherDiscussion})}.
Another limitation is related to the flexibility of synthesising controllers from a wide range of micro-controllers.
We have shown this in the context of few micro-controllers, and with a very simple meta-controller.
However, this limitation could be overcame by using a sophisticated controller such as  Rainbow~\cite{garlan2004} for the meta-controller.

Regarding external threats, a key issue that restricts the synthesis of controllers based on micro-controllers, as well as the reusability of such components, concerns the lack of repositories with existing micro-controllers. 
We highlight that a 
thorough evaluation of reusability can only be done by the community, as witnessed by Rainbow, whose initial evaluation of reusability was only
preliminary~\cite{garlan2004}, but 
% which was initially evaluated on just two target systems with different adaptation styles but same concern
eventually it was successfully reused in other pieces of work~\cite{yuan2013,schmerl2014,camara2013}.
%\fabiano{In what follows, we discuss our approach from a reusability perspective}.

\vspace{-0.1cm}

\subsection{Further Discussion}\label{sec:furtherDiscussion}

%\ap{[Rev.1] Rephrase in order to present microservices as a possible solution for implementing the approach}

The degree of controller reusability depends essentially on the micro-controllers. 
The principles of loose coupling and high cohesion associated with 
\fabiano{micro-controllers---analogously to}
microservices concepts \cite{Fowler2014}---promote reuse.
In terms of micro-controllers, this \fabiano{can be} achieved with service-specific micro-controllers with generic APIs.
% these micro-controllers could be stored in a repository.
Moreover, due to their simplicity, it can be easier to tailor existing micro-controllers towards the needs of a target system.
%, or indeed write completely a new one.
Additionally, micro-controllers that realise a closed loop promote flexibility and thus reuse. Based on feedback, a closed-loop micro-controller dynamically takes the needs of the specific target system into account (\eg, at the analysis stage of a controller, a closed-loop micro-controller for online testing can refine its testing strategy if the achieved test coverage of the target system is not sufficient). In contrast, an open-loop micro-controller requires predefined knowledge about the specific target system.  
This suggests that a closed-loop micro-controller has rather to be configured with generic techniques (\eg, to explore various testing strategies) while an open-loop micro-controller requires a target system-specific configuration (\eg, a testing strategy suited for the specific target system).

Besides individual micro-controllers, the coordination between them and how it is implemented also affect reuse.
%\ap{[Rev.1] In what follows, we can say that, the meta-controller processes adaptation rules for its target system (\ie the controller), similarly to what is done by the controller with respect to actual target system}
%\ap{We can rephrase. Composing a controller may be simpler (if few stages are needed), or complex (if there is uncertainty regarding the composition of the controller)}
The coordination is handled by the meta-controller and may follow a typical MAPE sequence, which may encompass a simple flow (if few stages are needed), or a complex flow (if there is uncertainty regarding the composition of the controller)

%As we mentioned in Section~\ref{sec:aprroach}, a micro-controller can be a closed loop and promote flexibility and reuse.  An example of a reusable micro-controller is the \textsf{FailureManager}. This micro-controller is responsible for identifying when/which sensors and actuators fail. As such, the data that the \textsf{FailureManager} generates could be useful for other applications and controllers that need to deal with sensor and actuator data. Figure~\ref{fig:reusingFailManager} illustrates a possible reuse of the \textsf{FailureManager} micro-controller in a controller that is also composed by \textsf{RegressionTesting} and \textsf{IntegrationTesting} micro-controllers which require failure-related data. 

%\begin{figure}[!ht]
%    \centering
%    \includegraphics[width=.9\columnwidth]{figures/reusingFailureManager.pdf}
%    \caption{A scenario for reusing a micro-controller. \ap{Remove this figure and associated text.}}
%    \label{fig:reusingFailManager}
%\end{figure}  

\section{Related Work} 
\label{sec:relatedWork}

We identified related work in two areas: the use of microservices for self-adaptive systems, and the reuse of controllers. We discuss both areas in the following.

\subsection{Microservices for Self-adaptive Systems}

%\important{Our work: novel approach on how controllers should be designed and deployed based on microservices, instead of discussing self-adaptation for target systems composed of microservices~\cite{mendonca2018}.}

\citet{Baylov2018} 
%\\(Reference Architecture for Self-adaptive Microservice Systems)
proposed a reference architecture to support the development of self-adaptive microservices. 
It requires a control loop inside each microservice, which share similarities with the micro-controllers we propose. However, adopting the reference architecture would imply a specific structure, which does not provide the same level of flexibility with respect to the composition of controllers with loosely coupled components (microservices) as our approach. 
%Thus, \citeauthor{Baylov2018} do not address the composition of a controller based on micro-controllers.

In the same context,
\citet{mendonca2018} 
%\\(Generality vs. Reusability in Architecture-based Self-adaptation: The Case for Self-adaptive Microservices)
discuss difficulties for reusing self-adaptation services and frameworks across different self-adaptive software systems.
From the generality and reusability perspectives, they argue that there is a mismatch between adaptation needs of modern systems and the current solutions.
Their proposed solution 
%-- not yet implemented --  
is to use containerised microservices as a primary abstraction to build the target system. 
The same research group have created Kubow~\cite{aberaldo2019}  
% \\(Kubow: An Architecture-Based Self-Adaptation Service for Cloud Native Applications)
as an extended and customised version of the Rainbow framework to provide self-adaptation support to containerised applications. 
Thus, the main focus is on realising self-adaptation for microservice-based target systems.
In a more recent report, 
\citet{mendonca2019} 
%\\(Developing Self-Adaptive Microservice Systems: Challenges and Directions)
discuss key challenges to build self-adaptive software systems based on microservices. One of them (challenge C5) concerns how to deploy controllers in this context. Regarding the granularity of the controller, two possibilities are mentioned but without a detailed solution: the controller being a single monolithic service, or being decomposed into a collection of independently developed and managed microservices.
%``\emph{How to determine the level of distribution, visibility and granularity necessary for deploying a microservice application's control components}''
\citet{Hassan2016} 
% \\(Microservices and Their Design Trade-Offs: A Self-Adaptive Roadmap)
proposed the creation of a controller for self-adaptive, microservice-based systems.
%Their focus is also on providing self-adaptation support for microservice-based target systems.
This is similar to what was proposed by
\citet{SampaioJr2019}.
%\\(Improving Microservice-based Applications with Runtime Placement Adaptation)
The latter \cite{SampaioJr2019} supports the reconfiguration of microservice-based systems according to affinities and history of resource usage of the composing microservices.

%\fcf{Highlighting the key differences with our work}
In contrast to  %\citeauthor{mendonca2018}'s, \citeauthor{aberaldo2019}'s, \citeauthor{Hassan2016}'s, and \citeauthor{SampaioJr2019}'s work,
the aforementioned approaches~\cite{Hassan2016, mendonca2018, aberaldo2019, SampaioJr2019}, our approach promotes the use micro-controllers (implemented as microservices) to architect flexible controllers.
%\fcf{Highlighting the "proximity" with our  work}
On the other hand, and aligned with \citeauthor{mendonca2019}'s suggestion~\cite{mendonca2019}, we propose and demonstrate a concrete solution for deploying adaptive, microservice-based controllers at run-time.

\citet{florio2016} 
%\\(Gru: An Approach to Introduce Decentralized Autonomic Behavior in Microservices Architectures)    
focused on adding  autonomic capabilities to containerised, microservice-based systems which were not originally designed to be autonomic. 
Their approach 
%named \emph{Gru}, 
consists of a
decentralised controller %(as a MAPE loop) 
implemented as a multi-agent system. %that is composed by a set of agents \todo{explain what an agent is}. 
Each agent realises a control loop 
that is responsible for managing a subset of microservices.
%The connection between the agents and the target system is realised through a virtual container---named \emph{autonomic enabler}---that wraps the target system and indirectly makes that target system autonomic.
\citet{Nallur2013} 
%\\(A Decentralized Self-adaptation Mechanism for Service-based Applications in the Cloud)
also developed a decentralised, multi-agent self-adaptation approach for web service-based systems deployed in the cloud.
Considering both pieces of work~\cite{florio2016, Nallur2013} and ours, key differences are:  
(i)~both \cite{florio2016, Nallur2013} were specifically designed to be applied to particular domains (namely, systems based on microservices, and systems based on web services, respectively); and
(ii) the controllers based on multi-agent systems are non-reconfigurable

% As as last note regarding microservices in the context of self-adaptive software systems, 
Finally, based on recent studies on microservice-based software systems~\citep{jamshidi2018,francesco2019} and on our discussion of related work, we were not able to identify any approach for composing and decomposing controllers from  micro-controllers to address the specific needs of individual target systems. This emphasises the originality of our work.

\begin{comment}

\fcf{We may not need to describe \citet{Silva2011}'s work 
%\\(Dynamic Plans for Integration Testing of Self-adaptive Software Systems)
(about integration testing) and \citet{Sharifloo2013}' work 
%\\(MCaaS: Model Checking in the Cloud for Assurances of Adaptive Systems)
(about model checking) here, since Section~\ref{sec:aprroach} already establishes a relationship with them.}
%\citet{Silva2011} proposed an approach to generate integration test plans during runtime. If errors are produced during the test execution, new plans are generated in order to enable more executions. 

%\citet{Silva2011} aimed to define test cases for the integration level at runtime, based on association with each component, identifying integration order, defining stubs and detecting error at runtime. The authors proposed an approach to generate test plans during runtime. First, test cases are associated with each available component. Second, there are mechanisms responsible for calculating the integration order of components. Third, all necessary stubs are also made available. Finally, in case of an error be detected during the execution of a generated plan, new plans are generated in order to enable more executions. 

\end{comment}

\subsection{Reuse for Self-adaptive Systems}

Related work in this area promotes reuse of various aspects of self-adaptive systems. % by using different approaches.
Examples are the MAPE-K blueprint~\cite{kephart2003}, design patterns~\cite{ramirez2010}, and control patterns~\cite{weyns2013} to support conceptual reuse of controllers.
% Reference models such as MAPE-K~\cite{Kephart:2003} describe architectural blueprints, \citet{Ramirez:2010} proposed design patterns, and \citet{Weyns:2013:patterns}  decentralised control patterns for self-adaptive systems. Such reference models and patterns promote reuse of controller solutions, however, only at the conceptual level.
%
Work supporting technical reuse raises the level of abstraction for self-adaptation.
Rainbow~\cite{garlan2004} raises the abstraction to the software architecture so that a controller performs generic architectural adaptation by adding, removing, and reconfiguring components in a target system. % ~\cite{Magee:1996}
For this purpose, Rainbow has to be tailored with target system-specific gauges and executors that bridge the abstraction gap, which
%
% More recently, with the rise of runtime models~\cite{Bennaceur:2014}, model-driven engineering techniques have been employed, ranging from bidirectional programming~\cite{Colson:2016} to incremental model synchronisation~\cite{Vogel:2010} to ease bridging the abstraction gap. 
could be eased by model-driven engineering techniques~\cite{Bennaceur:2014,vogel2010}. 
Abstracting from the target system, such approaches aim for generic but monolithic~controllers.

Other approaches enable reuse of execution engines for controllers specified by models~\cite{Iftikhar:2014,vogel2012}.
%, such as networks of timed automata~\cite{Iftikhar:2014} or with dedicated  languages~\cite{Vogel:2012}. 
The resulting models are specific for each target system but the engines executing these models are generic and reusable.
This principle has been extended to individual controller stages, which allows reuse of engines at a more fine-grained level~\cite{vogel2014}.
% (\eg, an OCL engine for the analysis stage)~\cite{Vogel2014-TAAS}.
While such a reuse eases developing controllers, the models have to be created for each specific target system. To ease the creation of models, reusable templates for each MAPE-K controller stage exist~\cite{DeLaIglesia2015}.
Similarly but for code-based development, \citet{krupitzer2015} provided reusable templates for components of these stages. % and the resulting components integrate with a reusable communication architecture.   
However, the controllers resulting from these approaches do not address the wide range of needs of target systems 
% (\eg, different levels of analysis depending on the system's criticality) 
since the unit of reuse is either a monolithic controller, templates typically restricting the structure or behaviour of controllers, or 
%implementation-level artefacts (e.g., execution engines).
model execution engines.
In contrast, we address the wide range of needs of target systems by synthesising a controller from a collection of generic micro-controllers.

\section{Conclusions} 
\label{sec:conclusions}

\rogerio{To promote the design of structurally  flexible  controllers for self-adaptive software systems, this paper proposes the synthesis of controllers from a collection of micro-controllers.}
%, which are service-specific microservices.
The management of this ensemble of micro-controllers is made by a meta-controller, which could be implemented by an existing monolithic controller, like Rainbow. 
\rogerio{The claim being made is that such a flexible controller structure, in the context of continuous deployment, would promote responsiveness to changes and reusability of micro-controllers across several application domains.} 
%\rdl{to discuss}
%Moreover, in the context of continuous deployment, controller design needs to be responsive to changes that affect the software of a target system.
The feasibility of the whole approach was demonstrated in the context of an Android case study.

It is clear that a two-tier controller tends to increase the complexity of a self-adaptive software system, and that may restrict the applicability of our solution.
For example, a solution based on micro-controllers would not be appropriate for applications that have limited amount of processing resources.
However, there is a whole range of self-adaptive software systems for which our proposal would be suitable, for example, systems that require diverse analytical and synthetical methods depending on their operational state, systems in which controller needs to be  dynamically reconfigured in response to changes, or systems that go through a wide range of changes that may affect either system services or their quality, including the provision of assurances. 

% \bento{With respect to operation state, involving as the target system as the controller, in this work we dealt with stateless micro-controllers. When it was necessary to use state in our case study, PhoneAdapter, the micro-controllers used shared data coming from the Knowledge component. Thus, as a future work, we intend to deal with issues related to stateful micro-controllers.}

The two-tier solution for controllers, however, also raises challenges that should be the basis for future work, and these include:
%establishing a repository of generic micro-controllers and quantitatively evaluating their reuse in different application domains,
\rogerio{testing whether an ensemble of micro-controllers behave as specified~\cite{hansel2015},}
assuring the switch between ensembles of micro-controllers~\cite{nahabedian2016}, especially when stateful micro-controllers are involved, and 
developing meta-controllers whose specific role is to manage micro-controllers~\cite{mendonca2018}.

%-----------------------------------------

\end{sloppypar}

%\clearpage
\linespread{1}

% \small % reduce the font size for References

\fontsize{8.5}{8.65} % to change the reference font size
\selectfont

\bibliographystyle{IEEEtranSN}
\bibliography{references}

%\clearpage
%\input{reviewsSEAMS2020}

\end{document}